# *Engineering Topological Phases with a Traveling-Wave Spacetime Modulation*

*João C. Serra*[*], *Mário G. Silveirinha*[†]

*University of Lisbon–Instituto Superior Técnico and Instituto de Telecomunicações, Avenida Rovisco Pais, 1, 1049-001 Lisboa, Portugal*

**Abstract**

Time-variant systems have recently garnered considerable attention due to their unique potentials in manipulating electromagnetic waves. Here, a novel class of topological spacetime crystals is introduced, with a traveling-wave modulation that mimics certain aspects of physical motion. Challenging intuition, our findings reveal that, even though such systems rely on a linear momentum bias, it is feasible to engineer an internal angular momentum and non-trivial topological phases by leveraging the symmetry of its structural elements. Furthermore, these platforms exhibit a gauge degree of freedom associated with the arbitrariness in the choice of the coordinate transformation that eliminates the time dependence of the system's Hamiltonian. The system's topology is intricately governed by a synthetic magnetic potential whose field lines can be controlled by manipulating material anisotropy. Remarkably, the proposed spacetime crystals host an unconventional class of scattering-immune edge states, whose oscillation frequency adapts continuously along the propagation path, shaped by the geometric attributes of the path itself.

---

* E-mail: joao.serra@lx.it.pt

† To whom correspondence should be addressed: mario.silveirinha@tecnico.ulisboa.pt

## 1. Introduction

Topological photonics provides an elegant theoretical framework to characterize the energy flow near the "boundaries" of material structures with complete band gaps [1-4]. Usually, topological systems are engineered by modulating the material response in space, but the crystal concept has been recently extended to structures presenting time-varying modulations [5-9]. This opens up many interesting possibilities as time and spacetime modulations enable exotic and nonreciprocal light-matter interactions without an external magnetic field bias, which can be useful for a myriad of optical applications [10-12]. Within this context, different topological aspects of time-dependent systems have been discussed in the literature [13-21].

Heuristically, nontrivial Chern phases are associated with some form of internal angular momentum [22, 23]. For instance, the nontrivial topology of a magnetized plasma is deeply rooted in the angular momentum generated by the static magnetic bias through the electron cyclotron orbits [24]. Accordingly, previous works have proposed different magnetless Chern insulators with spacetime modulations that implement a synthetic rotation [15, 21, 25]. The standard approach to modelling the intricate physics of these space-time-varying media and their resulting topological phases relies on effective descriptions, tight-binding approximations and perturbative analyses, which are only valid over a restricted region of the reciprocal wavevector-frequency space. However, topological invariants are global properties of the system that depend on all the eigenstates over the entire frequency spectrum, making local approximations prone to inaccuracies [26-29].

From a different perspective, these systems may be regarded as Floquet-Bloch insulators [15, 17, 20]. Importantly, as elaborated in the following section, the simultaneous space-time folding imposes a crucial constraint: the topological classification of such systems is only feasible if they have a finite spectral range (i.e., a bounded spectrum) in the absence of a time modulation. In contrast, photonic crystals exhibit an unbounded spectrum with an infinite number of bands, leading to a densely populated spectrum with no bandgaps when subjected to periodic time modulation.

In this article, we present the first strict topological classification of a photonic spacetime crystal by applying generalized Lorentz transformations to systems with a (linear) traveling-wave modulation, e.g., $\varepsilon = \varepsilon\left(\mathbf{r} - \mathbf{v}t\right)$ where $\mathbf{v}$ denotes the modulation velocity [30-33]. Counterintuitively, we show that one may engineer an angular momentum bias and nontrivial topological phases in crystals with a synthetic linear velocity. Remarkably, in this novel class of topological spacetime crystals, the bulk-edge correspondence ensures the existence of edge states that form robust waveguides, immune to back-scattering. Notably, the oscillation frequency of these states continuously adapts to the shape of the propagation path, potentially enabling frequency conversion without any back-scattering interference. This unique property may open opportunities for frequency-conversion mechanisms free of back-scattering interference.

It is worth noting that a previous work [35] has discussed related quantum space-time crystals in the context of a tight-binding model. However, the systems considered there lack a global traveling-wave spacetime symmetry, as one component of the Hamiltonian remains static. Thereby, they should be understood as generalizations of standard Floquet crystals.

## 2. Challenges in the Topological Classification of Space-Time Crystals

Our analysis is focused on platforms invariant to translations along the $z$-direction for which the transverse electric (TE, $H_z = 0$) and the transverse magnetic (TM, $E_z = 0$) polarizations are decoupled. Without loss of generality, we consider only TE polarized waves, fully described by the field components $\mathbf{\Psi} = \begin{pmatrix} E_z & H_x & H_y \end{pmatrix}^T$. In the absence of external sources, the system dynamics is governed by a Schrödinger-type equation:

$$\begin{pmatrix} 0 & -i\partial_y & +i\partial_x \\ -i\partial_y & 0 & 0 \\ +i\partial_x & 0 & 0 \end{pmatrix} \cdot \mathbf{\Psi}(\mathbf{r},t) = i\frac{\partial}{\partial t}\left[\mathbf{M}_{\text{TE}}(\mathbf{r},t)\cdot\mathbf{\Psi}(\mathbf{r},t)\right], \qquad \mathbf{M}_{\text{TE}}(\mathbf{r},t) = \begin{pmatrix} \varepsilon_{zz} & 0 & 0 \\ 0 & \mu_{xx} & \mu_{xy} \\ 0 & \mu_{xy} & \mu_{yy} \end{pmatrix}. \quad (1)$$

Note that TE-waves only probe the $zz$ component of the permittivity tensor and the in-plane components ($xx$, $xy$, $yy$) components of the (symmetric) permeability tensor. It will be shown later that materials exhibiting anisotropic responses play a crucial role in creating topological phases. Importantly, the material matrix $\mathbf{M}_{\text{TE}}(\mathbf{r},t)$ may vary simultaneously in space and time.

Let us begin by considering a regular photonic crystal with lattice vectors $\mathbf{a}_1$ and $\mathbf{a}_2$, i.e., the material matrix $\mathbf{M}_{\text{TE}}$ is time-independent and satisfies the discrete translation symmetries $\mathbf{M}_{\text{TE}}(\mathbf{r}) = \mathbf{M}_{\text{TE}}(\mathbf{r}+\mathbf{a}_i)$. In that case, the natural modes are Bloch waves of the type $\mathbf{\Psi}(\mathbf{r},t) = e^{-i\omega_{\mathbf{k}}t}e^{i\mathbf{k}\cdot\mathbf{r}}\mathbf{\Psi}_{\mathbf{k}}(\mathbf{r})$ with $\mathbf{\Psi}_{\mathbf{k}}(\mathbf{r}) = \mathbf{\Psi}_{\mathbf{k}}(\mathbf{r}+\mathbf{a}_i)$. As is well-known, the spatial periodicity of the Bloch envelope $\mathbf{\Psi}_{\mathbf{k}}(\mathbf{r})$ induces a periodicity in the Bloch wavevector space, i.e., $\omega_{\mathbf{k}} = \omega_{\mathbf{k}+\mathbf{b}_j}$ with $\mathbf{b}_j \cdot \mathbf{a}_i = 2\pi\delta_{ij}$, as represented in Fig. 1a for a 1D crystal.

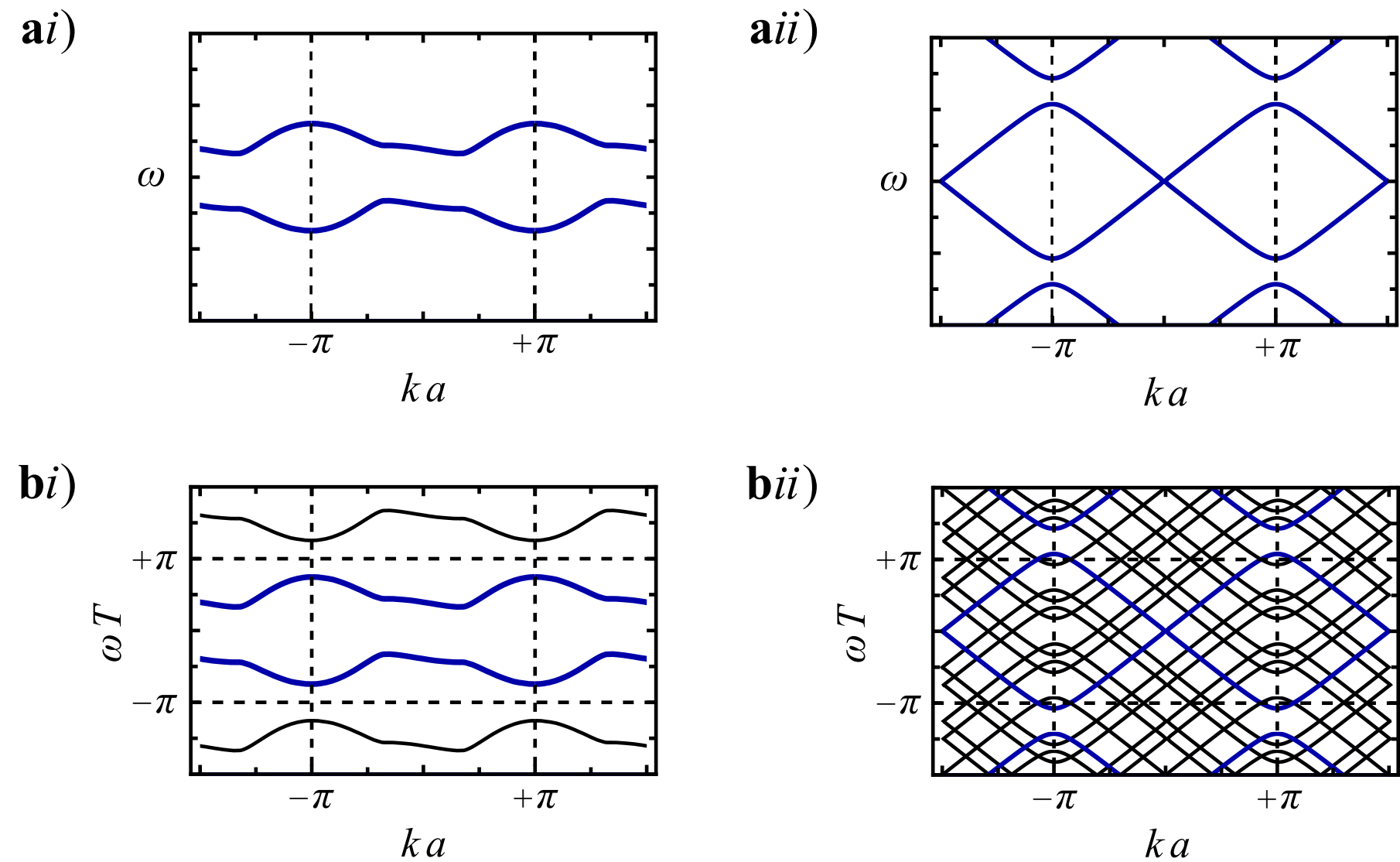

**Fig. 1 a)** Band structure of two spatially periodic systems: i) an array of single-mode resonators (discrete lattice model), with a finite number of photonic bands; ii) a photonic crystal with an infinite number of photonic bands. The spatial periodicity induces a Brillouin zone in reciprocal space, indicated by the vertical dashed lines. **b)** Band structures under an arbitrarily weak time modulation of the two crystals. The modulation causes the band structures to fold, resulting in periodicity along the frequency (vertical) axis. The horizontal dashed lines mark the unit cell in the frequency domain.

Now, suppose that the material response is periodically modulated in time with some period $T$. In that case, we get a space-time crystal, and the dispersion relation also becomes periodic along the Floquet frequency axis. For arbitrarily weak modulations, this equates to folding the time-independent band structure along the frequency axis with a spectral period $\Omega = 2\pi / T$ (empty lattice approximation) as illustrated in Fig. 1b.

When the original system is a lattice of single-mode resonators, the unperturbed band structure comprises two quasi-flat bands (Fig. 1ai), and weak modulations in time originate a discrete spectrum (free of accumulation points) with frequency gaps (Fig. 1bi) that can be topologically classified. However, photonic space crystals typically present an unbounded set of infinite positive and negative bands (Fig. 1aii), and the band-folding mechanism creates a very intricate dispersion without frequency gaps (Fig. 1bii), preventing a well-defined topological classification.

To circumvent this challenge, simplified models are often employed, such as tight-binding approximations (which describe the spacetime crystal as a lattice of resonators) [15, 25] or effective descriptions (which eliminate the space-time folding) [21]. However, these descriptions are only valid over a restricted region of the wavevector-frequency space and topological invariants are global properties of the system. Consequently, these methods have fundamental limitations, and general spacetime crystals lack a rigorous topological classification in the strict sense.

## 3. Traveling-Wave Space-Time Crystals

Notably, there exists a special class of systems where one can rigorously unfold the band structure and unveil hidden band gaps. These platforms are characterized by a material matrix $\mathbf{M}_{\mathrm{TE}}(\mathbf{r},t)$ with a traveling-wave type structure such that the space and time variations are synchronized as $\mathbf{M}_{\mathrm{TE}}(\mathbf{r},t)=\mathbf{M}_{\mathrm{TE}}(\mathbf{r}-\mathbf{v}t)$, where $\mathbf{v}$ is the modulation speed. This type of modulation shares some resemblances with moving systems [30-33], albeit the two in general are not equivalent [34].

The key idea for overcoming the challenges identified in the previous section is to apply a coordinate transformation that eliminates the explicit time dependence of the system. Due to the specific structure of the traveling-wave spacetime modulation, this transformation renders the system governed by a standard spatially periodic operator. As we shall show, this allows the application of conventional topological band theory, ensuring a well-defined topological classification.

Specifically, to characterize the spacetime crystal topology of such systems, we switch to a Lorentz co-moving frame where the material matrix $\mathbf{M}'_{\mathrm{TE}}$ becomes time independent. The electrodynamics in the co-moving frame (primed coordinates) is described by:

$$\mathbf{L}\left(-i\nabla'\right)\cdot\mathbf{\Psi}'\left(\mathbf{r}',t'\right)=i\mathbf{M}'_{\mathrm{TE}}\left(\mathbf{r}'\right)\cdot\frac{\partial}{\partial t'}\mathbf{\Psi}'\left(\mathbf{r}',t'\right),\qquad \mathbf{M}'_{\mathrm{TE}}=\begin{pmatrix}\varepsilon'_{zz} & \xi'_{zx} & \xi'_{zy}\\ \xi'_{zx} & \mu'_{xx} & \mu'_{xy}\\ \xi'_{zy} & \mu'_{xy} & \mu'_{yy}\end{pmatrix}. \tag{2}$$

The formulas for the transformed material parameters $\varepsilon'_{zz}$, $\mu'_{ij}$, $\xi'_{z,i}$ $(i,j=x,y)$ are given in the supplementary information [36]. In the above, $\mathbf{\Psi}'=\begin{pmatrix}E'_z & H'_x & H'_y\end{pmatrix}^T$ is the transformed state vector written in terms of the fields in the co-moving frame. The operator $\mathbf{L}\left(-i\nabla'\right)$ is the differential operator on the left-hand side of Eq. (1) with $\partial_x\to\partial_{x'}$ and $\partial_y\to\partial_{y'}$.

Importantly, the Lorentz transformation creates a magneto-electric coupling in the material response [32, 34, 37], here described by the elements $\xi'_{zx}$ and $\xi'_{zy}$, associated with a gauge degree of freedom due to the arbitrariness in the choice of the coordinate transformation. Indeed, there is a wide class of transformations that can be used to eliminate the time dependence of the material parameters, the simplest examples being the Lorentz and the Galilean transformations [34, 36]. For simplicity, we restrict our attention to the subluminal regime ($\left\|\mathbf{v}\right\|<v_{\mathrm{ph}}$ with $v_{\mathrm{ph}}$ the phase velocity of light in the background medium) to ensure that the system's spectrum remains real-valued, and all the studies in the main text rely on a standard Lorentz transformation.

In the supplementary information, we show that, in the non-relativistic limit, the state vector associated with a generic coordinate transformation (e.g., a Galilean transformation) is related to the state vector associated with the Lorentz transformation through a gauge transformation of the type $\mathbf{\Psi}' \rightarrow \mathbf{\Psi}' e^{i\mathbf{k}_\Delta \cdot \mathbf{r}'}$ with $\mathbf{k}_\Delta$ independent of the coordinates $\mathbf{r}'$ [36]. Thus, $\mathbf{\Psi}'$ may be understood as a gauge field.

Since the Lorentz transformation inherently mixes time and space, our coordinate transformation effectively alters the perceived geometry of the crystal. In particular, due to Lorentz contraction, the crystal appears spatially compressed in the laboratory frame compared to the co-moving frame, where the material response is purely space-dependent [34]. This effect is analogous to special relativity, where a moving object undergoes length contraction along the direction of motion. However, it is important to emphasize that in our case, there is no actual physical motion: the coordinate transformation is simply a mathematical tool that allows us to describe the system in a frame where its response is time-independent. Crucially, this transformation not only simplifies the analysis but also unveils the intrinsic topological properties of the original system, which are otherwise obscured in the laboratory frame due to the explicit time dependence of the modulation.

As already noted in previous works [21, 38, 39], there is a synthetic magnetic potential associated with the magneto-electric tensor. This synthetic potential plays a fundamental role in the description of the spacetime crystal, serving as the photonic counterpart of the standard vector potential in electromagnetism. Just as the conventional vector potential governs the behavior of charged particles in an electromagnetic field, the synthetic vector potential dictates how photons experience nonreciprocal effects in a modulated medium.

The dimensionless magnetic potential for TE waves is defined by $\mathbf{A}=\left[c\xi'_{zx}\hat{\mathbf{x}}+c\xi'_{zy}\hat{\mathbf{y}}\right]\times\hat{\mathbf{z}}$ [38]. While $\mathbf{A}$ depends on the gauge, i.e., it depends on the particular coordinate transformation, the corresponding synthetic magnetic field $\nabla\times\mathbf{A}$ is gauge independent [36]. This gauge invariance ensures that the physical effects arising from the modulation (e.g., the material topology) are independent of the specific coordinate transformation, i.e., of the gauge degree of freedom.

Importantly, as we shall detail in the following sections, the nonreciprocal effects in the system—and consequently, the topological properties of the photonic crystal—are effectively governed by this synthetic vector potential. In analogy with condensed matter systems, where an external magnetic field can induce topologically nontrivial phases in electronic band structures, here the synthetic vector potential provides a convenient and intuitive framework for understanding how time-modulated systems acquire topological properties.

## 4. Traveling-Wave Modulation of a Honeycomb Lattice

To begin with, let us consider a spacetime crystal formed by isotropic materials ( $\varepsilon_{zz}\equiv\varepsilon$, $\mu_{xx}=\mu_{yy}\equiv\mu$ and $\mu_{xy}=0$). In this case, the direction of the synthetic potential is strictly locked to the direction of the modulation velocity as $\mathbf{A}\approx-\left(n^2-1\right)\mathbf{v}/c$. Here, $n=c\sqrt{\varepsilon\mu}$ is the refractive index of the material, and for simplicity we ignore second order relativistic corrections [36]. The most intuitive way to generate internal angular momentum is by engineering a magnetic potential $\mathbf{A}$ with an azimuthal-type structure, ensuring that the vector field lines follow circular orbits. This solution was studied in our previous work [21] using an effective medium formalism and rotating spacetime modulations. However, the linear spacetime modulation is seemingly incompatible with such solutions, due to the aforementioned restriction on the direction of $\mathbf{A}$, which supports the heuristic reasoning that a linear momentum bias is unsuitable to generate topological Chern phases.

To validate this argument, we consider a spacetime crystal formed by cylindrical isotropic rods (Fig. 2a). Due to the Lorentz contraction, the geometry of the spacetime crystal is slightly different in the laboratory and co-moving frames [34]. For simplicity, we fix the lattice structure (honeycomb lattice) and the geometrical shapes (circular rods) in the co-moving frame. In the laboratory frame coordinates, all the geometrical distances are contracted by the Lorentz factor $\gamma = \left(1 - v^2 / c^2\right)^{-1/2}$ along the direction of $\mathbf{v}$ and the rods have slightly ellipsoidal cross-sections. There are two inequivalent rods per unit cell (represented in red and green in Fig. 2a) with permittivities $\varepsilon_1$, $\varepsilon_2$ and a trivial permeability in the laboratory frame. The background region is taken as air, so that its response is unaffected by the Lorentz transformation [34].

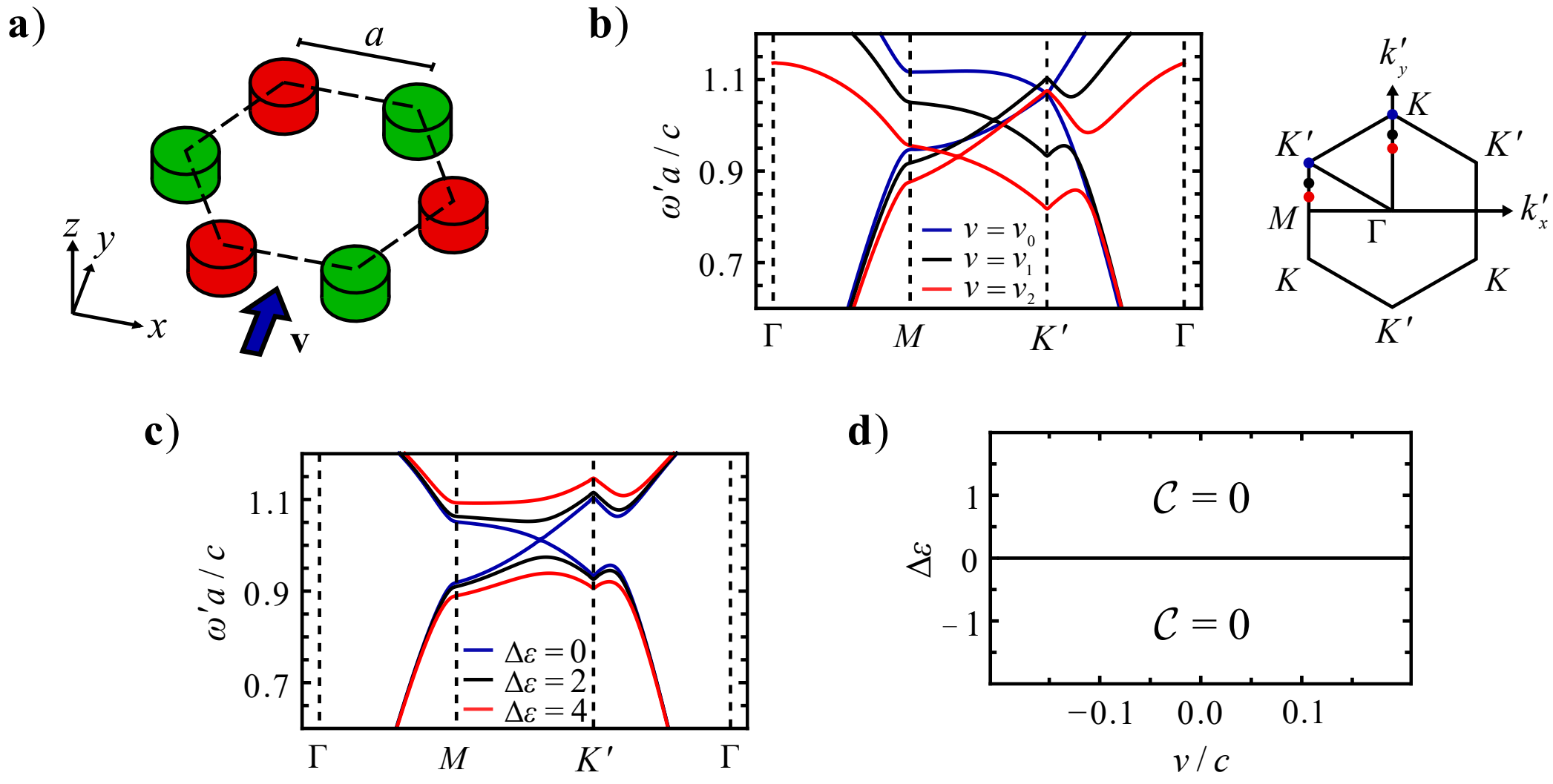

**Fig. 2** **a)** Unit cell of a spacetime crystal with modulation velocity $\mathbf{v} = v\hat{\mathbf{y}}$. The unit cell contains two dielectric inclusions with permittivity $\varepsilon_1$ (red) and $\varepsilon_2$ (green). In the Lorentz co-moving frame, the crystal has a honeycomb lattice and the inclusions have a circular cross-section with radius $R = 0.3a$, being $a$ the nearest neighbors distance. In the laboratory frame, the crystal is Lorentz contracted. **b)** Band structure in the Lorentz co-moving frame of a crystal with identical inclusions ($\varepsilon_1 = \varepsilon_2 = 12\varepsilon_0$) for different modulation speeds: $v_0 = 0.0c$ (blue), $v_1 = -0.1c$ (black), $v_2 = -0.2c$ (red). The Dirac degeneracies are not lifted by the synthetic motion, instead their positions are shifted in the Brillouin Zone along the direction of $\mathbf{v}$ (right hand-side inset). **c)** Band structure in the Lorentz co-moving frame with a modulation speed $v = -0.1c$ for inclusions with $\varepsilon_1 = \left(12 + \frac{\Delta\varepsilon}{2}\right)\varepsilon_0$ and $\varepsilon_2 = \left(12 - \frac{\Delta\varepsilon}{2}\right)\varepsilon_0$ for $\Delta\varepsilon = 0$ (blue line), $\Delta\varepsilon = 2$ (black line), $\Delta\varepsilon = 4$ (red line). **d)** Topological phase diagram for different values of the permittivity detuning $\Delta\varepsilon$ and of the modulation speed $v$. If the $\mathcal{P}\cdot\mathcal{T}$ symmetry is preserved ($\Delta\varepsilon = 0$), the band gap remains closed; otherwise, a complete band gap is formed but with a trivial topology.

The band structure of a honeycomb lattice is characterized by Dirac cones at the high-symmetry points $K$ and $K'$ [40-42]. In most systems, breaking either the parity ($\mathcal{P}$) or time-reversal ($\mathcal{T}$) symmetries is sufficient to lift this degeneracy (here, the operator $\mathcal{P}:(x,y)\rightarrow(-x,-y)$ represents a two-fold rotation). As time-varying systems have a broken $\mathcal{T}$ symmetry, the modulation of the material parameters can potentially open a topologically nontrivial gap. Figure 2b presents the band structure of the spacetime crystal in the Lorentz co-moving frame calculated using the plane wave expansion method for different modulation speeds along the $y$-axis [36]. The rods are assumed identical ($\varepsilon_1 = \varepsilon_2 = 12\varepsilon_0$). Surprisingly, the Dirac degeneracies are not lifted for $v \neq 0$. Instead, they are shifted in the momentum space $\left(k'_x, k'_y\right)$ along the direction of $\mathbf{v}$. This result holds true for an arbitrary orientation and amplitude of $\mathbf{v}$. Importantly, even if the band gap could be opened, it would necessarily be topologically trivial. In fact, when $\varepsilon(\mathbf{r},t=0)=\varepsilon(-\mathbf{r},t=0)$ the system is $\mathcal{P}\cdot\mathcal{T}$ symmetric and the Berry curvature vanishes, resulting in a trivial topology [36]. Note that both the time-reversal and the parity operators flip the direction of the modulation velocity, so $\mathbf{v}$ remains unchanged under the composition of the two operators.

From the previous discussion, it is manifest that a nontrivial topology requires breaking the two-fold rotation (parity) symmetry of the static crystal, i.e., guarantee that $\varepsilon(\mathbf{r},t=0)\neq\varepsilon(-\mathbf{r},t=0)$. The simplest way to do this is to detune the permittivity of the two inclusions so that $\varepsilon_{1,2}/\varepsilon_0 = 12 \pm \Delta\varepsilon/2$. As shown in Fig. 2c, a permittivity detuning $\Delta\varepsilon \neq 0$ does indeed create a full gap in the co-moving frame band diagram, but the topological charge is always trivial. This is consistent with well-known properties of the Haldane model, where the parity symmetry breaking always favours trivial topologies [43]. A general discussion on how symmetry constraints the formation of topological phases in spacetime crystals can be found in the supplementary information [36].

Figure 2d summarizes our findings in the form of a topological phase diagram. The gap Chern number is numerically calculated using the Green's function method [44-47] (see supplementary information [36]). Note that the phase diagram is precisely the same for other spacetime coordinate transformations (e.g., for a Galilean transformation) because the eigenstates obtained with different gauges are related as $\mathbf{\Psi}' \to \mathbf{\Psi}' e^{i\mathbf{k}_\Delta \cdot \mathbf{r}'}$.

## 5. Using Anisotropy to Engineer an Internal Angular Momentum

As previously mentioned, for isotropic spacetime crystals the direction of the synthetic vector potential is locked to the modulation speed as $\mathbf{A} \approx -\left(n^2-1\right)\mathbf{v}/c$. Interestingly, for an isorefractive crystal formed by isotropic materials with $n=1$ the vector potential vanishes exactly. In fact, materials with $n=1$, denoted by Minkowskian isorefractive media, are "fixed points" of the Lorentz transformation, i.e., such media have the same constitutive relations in any inertial frame [34, 48-50]. Crystals formed by Minkowskian isorefractive media have a trivial magneto-electric response in the Lorentz co-moving frame ($\xi'_{zx} = \xi'_{zy} = 0$) and hence are always topologically trivial.

It is useful to analyze how the vector potential $\mathbf{A}$ changes when the condition $n = const.$ is slightly perturbed with a weak magnetic anisotropy. For conciseness here, we restrict the discussion to the Minkowskian case ($n=1$), but all the results can be readily extended to isorefractive crystals formed by isotropic materials with $1 < n = const.$ [36]. To this end, we consider a spacetime crystal such that the in-plane permeability is of the form $\mu\left[\left(1+\delta_\mu\right)\hat{\mathbf{e}}_1\otimes\hat{\mathbf{e}}_1+\left(1-\delta_\mu\right)\hat{\mathbf{e}}_2\otimes\hat{\mathbf{e}}_2\right]$. Here, $\hat{\mathbf{e}}_1$, $\hat{\mathbf{e}}_2$ are the in-plane orthogonal permeability main axes, defined so that $\hat{\mathbf{e}}_1\times\hat{\mathbf{e}}_2=\hat{\mathbf{z}}$. The in-plane permeability eigenvalues are $\mu\left(1\pm\delta_\mu\right)$. All the parameters ($\hat{\mathbf{e}}_1$, $\hat{\mathbf{e}}_2$, $\mu$, $\delta_\mu$) are functions of $\mathbf{r}-\mathbf{v}t$ in the laboratory frame coordinates. Furthermore, it is supposed that $\varepsilon_{zz}\mu=1/c^2$ so that $\delta_\mu$ gives the detuning with respect to the isorefractive case ($\delta_\mu=0$). Such an anisotropic crystal implements a synthetic vector potential given by [36]:

$$\mathbf{A}=\frac{v}{c}\delta_\mu\hat{\mathbf{e}}_{\mathrm{A}}, \quad \text{with} \quad \hat{\mathbf{e}}_{\mathrm{A}}\equiv\cos\left(2\varphi_\mu-\varphi_v\right)\hat{\mathbf{x}}+\sin\left(2\varphi_\mu-\varphi_v\right)\hat{\mathbf{y}}. \tag{3}$$

As seen, the vector potential is proportional to both the velocity $v$ and to the permeability detuning $\delta_\mu$, so that $\mathbf{A}$ vanishes in the Minkowskian case ($n=1$). In the above, $\varphi_\mu$ ($\varphi_v$) denotes the angle of the main axis $\hat{\mathbf{e}}_1$ (modulation speed $\mathbf{v}$) with respect to the $x$-direction (Fig. 3a). Remarkably, the direction of the vector potential is not locked to the modulation speed, but depends as well on the axes of the permeability tensor. In particular, by changing continuously the direction ($\varphi_\mu$) of $\hat{\mathbf{e}}_1$, it is possible to engineer a vector potential with an arbitrary orientation! This property is illustrated in Fig. 3a: the direction of $\mathbf{A}$ is obtained through a rotation of the vector $\mathbf{v}$ by an angle of $\Delta\varphi=2\left(\varphi_\mu-\varphi_v\right)$ with respect to the $z$-axis.

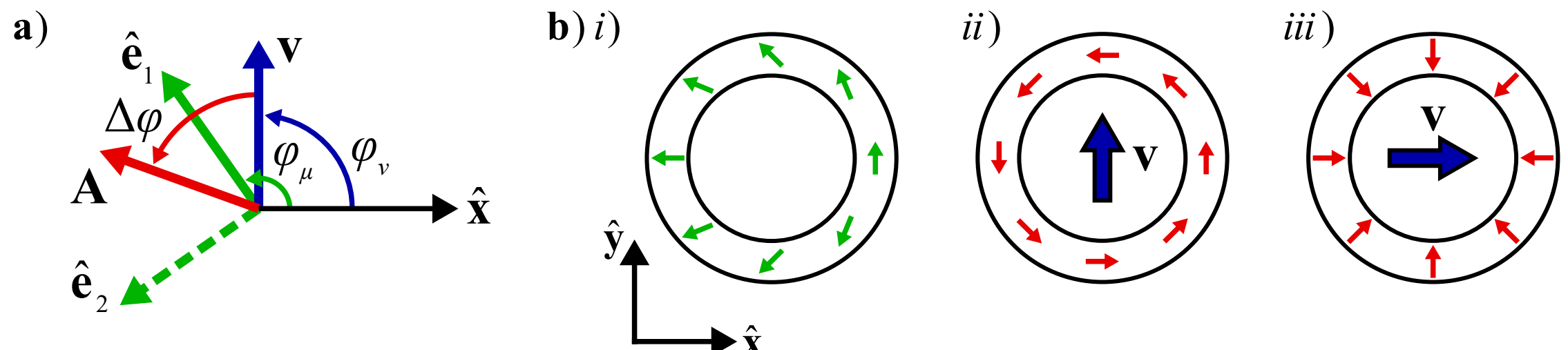

**Fig. 3 a)** Geometrical relation between the modulation velocity (blue), the permeability main axes (green) and the synthetic magnetic potential (red). The orientation of the modulation velocity and the magnetic potential differ by the angle $\Delta\varphi = 2\left(\varphi_\mu - \varphi_v\right)$. The material response is a slight anisotropic perturbation of a Minkowskian isotropic medium. **b**i) Geometry of an anisotropic ring such that the permeability main axes rotate continuously by a half-turn along its perimeter. The direction of the main axis $\hat{\mathbf{e}}_1$ is represented by the green arrows. The synthetic magnetic potential $\mathbf{A}$ field lines are represented by the red arrows for $\mathbf{v} = v\hat{\mathbf{y}}$ in (**b**ii) and for $\mathbf{v} = v\hat{\mathbf{x}}$ in (**b**iii). The synthetic motion induces an internal angular momentum only in ii).

Motivated by this result, we introduce an anisotropic ring (Fig. 3bi) which, when subjected to a linear modulation with $\mathbf{v} = v\hat{\mathbf{y}}$, implements a vector potential with an azimuthal structure in the Lorentz comoving frame. The ring is formed by an anisotropic material with the main axis $\hat{\mathbf{e}}_1$ of the permeability tensor (green arrows) varying continuously along the ring perimeter such that $\varphi_\mu = \pi/2 + \varphi/2$ with $\varphi$ the polar angle measured with respect to the ring centre. Note that in a full cycle $\hat{\mathbf{e}}_1$ is transformed into $-\hat{\mathbf{e}}_1$, but the permeability tensor is transformed as $\boldsymbol{\mu} \to +\boldsymbol{\mu}$ because it is insensitive to the "polarity" of $\hat{\mathbf{e}}_1$. This property is reminiscent of the "fermionic half-turn" periodicity of the polarization vector in a conical diffraction ring [51]. Importantly, the ring has a broken two-fold rotation symmetry. Figures 3bii and 3biii show the vector field lines of the synthetic magnetic potential [Eq. (3)] when the modulation speed is directed along the $y$ and $x$ directions, respectively. Remarkably, for $\mathbf{v} = v\hat{\mathbf{y}}$ the $\mathbf{A}$ lines are azimuthal, whereas for $\mathbf{v} = v\hat{\mathbf{x}}$ they are radial.

A spacetime crystal formed by many of such rings (see Sec. 6) has a nontrivial internal angular momentum (heuristically defined as $\mathcal{L} \sim \int ds' \, \hat{\mathbf{z}} \cdot \left( \mathbf{r}' \times \mathbf{A} \right)$) and a nontrivial topology only when $\mathbf{v} = v\hat{\mathbf{y}}$. In fact, as further discussed in the supplementary information [36], a spacetime crystal with a mirror plane parallel to the modulation velocity is necessarily topologically trivial. Since the anisotropic ring has a mirror plane at $y = 0$, $\mathbf{v} = v\hat{\mathbf{x}}$ leads to a trivial topology. Gauge fields relying on anisotropic materials were previously discussed in a different context [52].

## 6. Nontrivial Topological Phases and Bulk-Edge Correspondence

Figure 4a depicts a spacetime crystal constructed from an isorefractive system (elements in red plus the background region) with a refractive index $n = 1$ (Minkowskian media) perturbed with a discrete version of the anisotropic ring (elements in gray). Thereby, the synthetic magnetic potential is generated exclusively by the elements in the center. The role of the red elements is to engineer band degeneracies at the Dirac points that are lifted by the spacetime modulation.

The 3 elements in gray are placed along the perimeter of a circle with radius $R = 0.35a$ at the azimuth angles $\varphi_n = -\frac{\pi}{2} + \left(n-1\right)\frac{2\pi}{N}$ with $n = 1,...,N$ and $N = 3$.The scatterers are formed by an anisotropic material with $\delta_\mu = 0.5$, with the main axis $\hat{\mathbf{e}}_1$ of the permeability tensor directed along $\varphi_{\mu,n} = \frac{\pi}{2} + \frac{\varphi_n}{2}$. The generated synthetic magnetic potential is shown in Fig. 4a.

Figure 4b depicts the band structure in the Lorentz co-moving frame for $v = 0.2c$. The band gap shaded in blue is characterized by $\mathcal{C}_{\text{gap}} = -1$, in agreement with the azimuthal-type magnetic potential field lines in Fig. 4a. The corresponding Berry curvature is mostly concentrated at the two Dirac points [36]. In the supplementary materials, we present a detailed convergence study of the gap Chern number. We wish to underline that the numerical results in Fig. 4 fully account for the relativistic-type effects arising from the coordinate transformation.

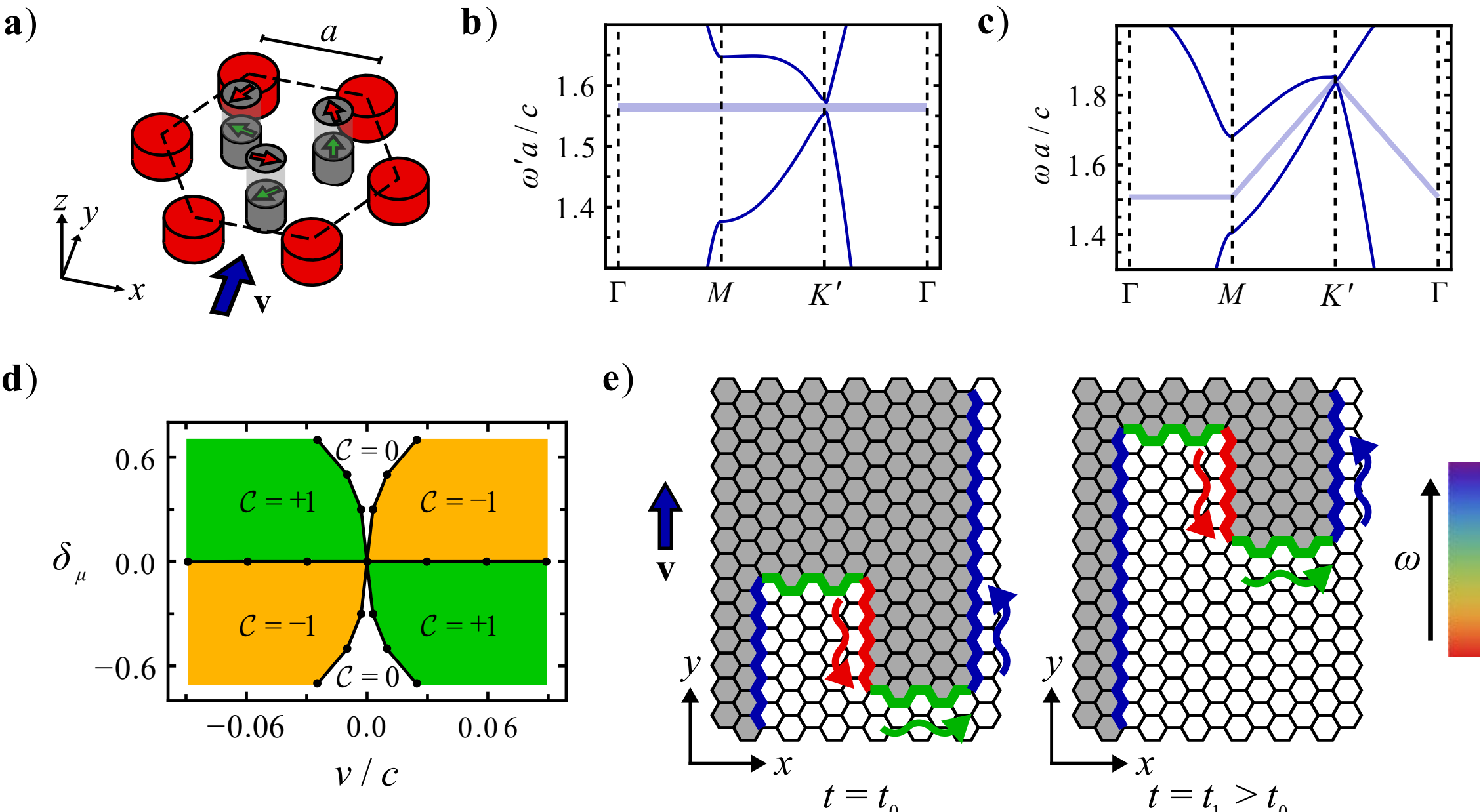

**Fig. 4 a)** Unit cell of a spacetime crystal constructed from an isorefractive system (air background plus the honeycomb array of isotropic red elements with $\varepsilon = 10\varepsilon_0$ and $\mu = 0.1\mu_0$) perturbed with the gray anisotropic elements. The gray elements provide a discrete realization of the anisotropic ring (Fig. 3bi). They have an anisotropic response characterized by the detuning parameter $\delta_\mu = 0.5$ and the permeability main axis $\hat{\mathbf{e}}_1$ is represented by the green arrows. We represent the corresponding synthetic magnetic potential $\mathbf{A}$ in the Lorentz co-moving frame with the red arrows above the cylinders. The red and gray elements have circular cross-sections with radii $R_r = 0.2a$ and $R_g = 0.17a$, respectively. All the dimensions are specified in the Lorentz co-moving frame. **b)** Band structure in the Lorentz co-moving frame for a modulation speed $v = 0.2c$ along the $y$-direction. The band gap is characterized by $\mathcal{C}_{\text{gap}} = -1$. **c)** Band structure in the laboratory frame. **d)** Topological phase diagram as a function of the modulation speed $v$ and of the detuning parameter $\delta_\mu$. **e)** Time evolution of the interface between two topologically inequivalent spacetime crystals (shaded in gray and in white colors), as seen in the lab frame. Both crystals are subject to the same modulation speed $\mathbf{v}$ along the $y$-direction. It is supposed that the interface supports a single unidirectional edge state. Due to the synthetic Doppler shift, the frequency observed in the lab frame changes with the direction of propagation of the edge state (see the color bar).

The band structure in the laboratory frame can be found from the band structure in the co-moving frame using a relativistic Doppler transformation $k_x = k'_x$, $k_y = \gamma\left(k'_y + \omega' v / c^2\right)$ and $\omega = \gamma\left(\omega' + v k'_y\right)$ [34, 53]. Consequently, as shown in Fig. 4c, the band gap is no longer a horizontal strip in the $(\mathbf{k}, \omega)$-space, but rather a slanted strip along the $k_y$ direction in the laboratory frame coordinates.

Figure 4d shows the topological phase diagram in terms of the modulation speed $v$ and the detuning parameter $\delta_\mu$. A nontrivial topology is feasible only when $v$ is larger than some threshold value that depends on $\delta_\mu$. Similar to the synthetic vector potential [Eq. (3)], the gap Chern number changes sign when either $v$ or $\delta_\mu$ change sign. In our system, the Chern number is restricted to the values $-1$, $0$ and $+1$, but in principle one can also implement higher-order Chern numbers by resorting to spacetime crystals with “supercell modulations” as in Ref. [54]. In the supplementary materials, we present a study of other implementations of the anisotropic ring [36]. It is shown that a ring with a larger number of elements provides more robust topological phases. Furthermore, we also prove that it is possible to engineer nontrivial topologies using only isotropic scatterers [36].

The bulk-edge correspondence principle holds rigorously in the co-moving frame, provided that the boundary walls of the system remain time-independent in this frame. Under this condition, the number of topological edge states is dictated by the gap Chern number, ensuring the existence of edge modes that propagate without back-scattering. Since the co-moving and laboratory frames describe the same physical system, this result also directly applies in the laboratory frame. Specifically, the number of edge states in the lab frame remains constrained by the gap Chern number and immune to back-scattering.

Importantly, in the lab frame, the frequency of oscillation of the edge states may vary along the propagation path due to a synthetic Doppler shift that arises from the Lorentz transformation. For an edge mode propagating along the direction $\hat{\mathbf{t}}'$ (tangent to the boundary wall), its frequency as observed in the laboratory frame is $\omega = \gamma\left(\omega' + k'_{\text{egde}}\hat{\mathbf{t}}' \cdot \mathbf{v}\right)$. Here, $\omega'$ and $k'_{\text{egde}}$ are the frequency and propagation constant of the edge mode in the co-moving frame coordinates. Note that $k'_{\text{egde}}$ may depend on $\hat{\mathbf{t}}'$, but $\omega'$ is independent of the propagation path. As the interface between two topologically inequivalent spacetime crystals undergoes deformation, the frequency of the primary oscillation shifts accordingly, as sketched in Fig. 4e. This property enables waves to travel along a spacetime interface without any form of back-scattering, all while its frequency varies along the propagation path to conserve the quantity $\omega - \mathbf{v} \cdot \mathbf{k}_{\text{edge}}$. This feature opens up intriguing possibilities for employing topological spacetime crystals in frequency-conversion applications, while entirely avoiding any type of back-scattering interference.

Although our current designs rely on materials with both electric and magnetic responses, in principle it should be possible to devise related spacetime crystals based solely on a permittivity response. This might pave the way for realistic implementations, for example based on microwave technologies and time modulated varactors. Note that in a practical realization, the edge states can be excited and probed with stationary antennas. Traveling-wave antennas may be particularly effective in ensuring the excitation of a single edge state.

## 7. Conclusions

In summary, we introduced a novel class of topological systems relying on spacetime traveling-wave modulations. We have demonstrated that, despite the linear momentum bias inherent in such systems, it is indeed possible to engineer an internal angular momentum and harness topological phases by leveraging the anisotropy of the materials. Notably, we have established that spacetime crystals possess a gauge degree of freedom tied to the selection of the coordinate transformation that eliminates the time dependence of the material response. The system topology is dictated by a gauge-dependent synthetic magnetic field whose orientation can be controlled by tailoring the main axes of the material tensors. The introduced systems open up an exciting new paradigm for light transport that defies back-scattering with the frequency of edge modes controlled by the shape of the spacetime boundary.

## Acknowledgements

This work is supported in part by the Institution of Engineering and Technology (IET), by the Simons Foundation Award SFI-MPS-EWP-00008530-10, and by FCT/MECI through national funds and when applicable co-funded EU funds under UID/50008: Instituto de Telecomunicações.

## Conflict of Interest

The authors declare no conflict of interests.

# Supplementary Information:

## *“Engineering Topological Phases with a Traveling-Wave Spacetime Modulation”*

*João C. Serra*[*], *Mário G. Silveirinha*[†]

University of Lisbon–Instituto Superior Técnico and Instituto de Telecomunicações, Avenida Rovisco Pais, 1, 1049-001 Lisboa, Portugal

The supplementary information provides additional details on A) generalized Lorentz transformations, B) gauge transformations, C) the synthetic magnetic potential, D) the numerical calculation of the band structure of the spacetime crystal, E) the numerical calculation of the gap Chern number, F) symmetry constraints on the emergence of unidirectional edge modes, G) spacetime crystals with a square lattice, H) other implementations of the “anisotropic ring”.

### *A. Generalized Lorentz transformations*

In this supplementary note, we discuss how the material response changes under a generalized Lorentz transformation. Consider a linear time-variant material described by the laboratory frame constitutive relations

$$\begin{pmatrix} \mathbf{D}(\mathbf{r},t) \\ \mathbf{B}(\mathbf{r},t) \end{pmatrix} = \underbrace{\begin{pmatrix} \boldsymbol{\varepsilon}(\mathbf{r}-\mathbf{v}t) & \mathbf{0}_{3\times3} \\ \mathbf{0}_{3\times3} & \boldsymbol{\mu}(\mathbf{r}-\mathbf{v}t) \end{pmatrix}}_{\mathbf{M}} \begin{pmatrix} \mathbf{E}(\mathbf{r},t) \\ \mathbf{H}(\mathbf{r},t) \end{pmatrix}, \tag{S1}$$

* E-mail: joao.serra@lx.it.pt

† To whom correspondence should be addressed: mario.silveirinha@tecnico.ulisboa.pt

with $\boldsymbol{\varepsilon}$ and $\boldsymbol{\mu}$ symmetric tensors. We introduce the generalized Lorentz transformation of coordinates [16, 27]:

$$\mathbf{r}' = \left(\mathbf{1}_{\perp} + \gamma \hat{\mathbf{u}}_{\parallel} \otimes \hat{\mathbf{u}}_{\parallel}\right) \cdot \mathbf{r} - \gamma \mathbf{v} t, \qquad t' = \gamma\left(t - \frac{\mathbf{v}}{c_0^2} \cdot \mathbf{r}\right) \tag{S2}$$

Here, $\mathbf{v} = v\hat{\mathbf{u}}_{\parallel}$, $\mathbf{1}_{\perp} = \mathbf{1}_{3\times 3} - \hat{\mathbf{u}}_{\parallel} \otimes \hat{\mathbf{u}}_{\parallel}$ and $\gamma = 1/\sqrt{1-\left(v/c_0\right)^2}$. For a regular Lorentz transformation, $c_0 = c$ is the speed of light in vacuum. In some situations, it is useful to consider other generalized Lorentz transformations with $0 < c_0 \leq \infty$ (see Ref. [27]). For example, the case $c_0 = \infty$ corresponds to a Galilean transformation of coordinates which is widely used to characterize the electromagnetic response of spacetime crystals with a traveling-wave modulation [15, 16, 27]. Thus, there is a wide family of transformations (specifically those with $c_0 > v$) that can be used to eliminate the time dependence of the material response. As further discussed in the supplementary note B, this arbitrariness in the choice of the coordinate transformation can be regarded as a gauge degree of freedom that exists in both the subluminal and superluminal regimes.

Following Refs. [16, 27], the transformed electromagnetic fields are linked in the new coordinates as

$$\begin{pmatrix} \mathbf{D}'(\mathbf{r}',t') \\ \mathbf{B}'(\mathbf{r}',t') \end{pmatrix} = \mathbf{M}' \cdot \begin{pmatrix} \mathbf{E}'(\mathbf{r}',t') \\ \mathbf{H}'(\mathbf{r}',t') \end{pmatrix} \tag{S3}$$

with $\mathbf{M}'$ the transformed (time-independent) $6\times 6$ material matrix given by:

$$\mathbf{M}' = \left[\frac{1}{c_0^2}\mathbf{V} + \mathbf{A} \cdot \mathbf{M}\right] \cdot \left[\mathbf{A} + \mathbf{V} \cdot \mathbf{M}\right]^{-1}, \quad \text{with} \tag{S4a}$$

$$\mathbf{V} = \begin{pmatrix} 0 & \gamma \mathbf{v}\times\mathbf{1} \\ -\gamma \mathbf{v}\times\mathbf{1} & 0 \end{pmatrix}, \qquad \mathbf{A} = \begin{pmatrix} \gamma \mathbf{1}_\perp + \hat{\mathbf{u}}_\parallel \otimes \hat{\mathbf{u}}_\parallel & 0 \\ 0 & \gamma \mathbf{1}_\perp + \hat{\mathbf{u}}_\parallel \otimes \hat{\mathbf{u}}_\parallel \end{pmatrix}. \tag{S4b}$$

The exact definition of the transformed fields can be found in Ref. [27]. The transformed fields satisfy Maxwell's equations:

$$\begin{pmatrix} 0 & i\nabla'\times\mathbf{1} \\ -i\nabla'\times\mathbf{1} & 0 \end{pmatrix} \cdot \begin{pmatrix} \mathbf{E}'(\mathbf{r}',t') \\ \mathbf{H}'(\mathbf{r}',t') \end{pmatrix} = i\mathbf{M}' \cdot \frac{\partial}{\partial t'} \begin{pmatrix} \mathbf{E}'(\mathbf{r}',t') \\ \mathbf{H}'(\mathbf{r}',t') \end{pmatrix}. \tag{S5}$$

It is convenient to write the transformed material matrix as:

$$\mathbf{M}' = \begin{pmatrix} \boldsymbol{\varepsilon}' & \boldsymbol{\xi}' \\ \boldsymbol{\zeta}' & \boldsymbol{\mu}' \end{pmatrix}, \tag{S6}$$

where $\boldsymbol{\varepsilon}'$ is the permittivity, $\boldsymbol{\mu}'$ is the permeability and $\boldsymbol{\xi}'$, $\boldsymbol{\zeta}'$ are the magneto-electric tensors in the co-moving frame coordinates.

Let us consider now a scenario where the permittivity and permeability tensors are of the form:

$$\boldsymbol{\varepsilon} = \begin{pmatrix} \varepsilon_{xx} & \varepsilon_{xy} & 0 \\ \varepsilon_{xy} & \varepsilon_{yy} & 0 \\ 0 & 0 & \varepsilon_{zz} \end{pmatrix}, \qquad \boldsymbol{\mu} = \begin{pmatrix} \mu_{xx} & \mu_{xy} & 0 \\ \mu_{xy} & \mu_{yy} & 0 \\ 0 & 0 & \mu_{zz} \end{pmatrix}, \tag{S7}$$

so that the TE and TM polarizations can be decoupled into two sets of independent equations. Then, it can be easily checked that Maxwell's equations [Eq. (S5)] for TE-waves reduce to Eq. (2) of the main text. The elements $\varepsilon'_{zz}$, $\mu'_{xx}$, $\mu'_{yy}$, $\mu'_{xy}$, $\xi'_{zx}$, $\xi'_{zy}$ can be found by inserting Eq. (S7) into Eqs. (S4) and (S6), as detailed in the following subsections. The dynamics for TM-polarized waves is related to Eq. (1) from the main text through a simple duality transformation. From a practical standpoint, choosing a specific field polarization might offer advantages and simplicity in terms of implementation, depending on the solution that is exploited to modulate the electric and magnetic responses of the materials.

*I. Non-relativistic limit*

Let us first ignore the relativistic corrections, i.e., consider a first order Taylor expansion of $\mathbf{M}'$ in the modulation speed. Then, using $\mathbf{r}' \approx \mathbf{r} - \mathbf{v}t$, $\mathbf{V} \approx \begin{pmatrix} 0 & \mathbf{v}\times\mathbf{1} \\ -\mathbf{v}\times\mathbf{1} & 0 \end{pmatrix}$, $\mathbf{A} \approx \mathbf{1}_{6\times6}$, and $\left[\mathbf{A} + \mathbf{V}\cdot\mathbf{M}\right]^{-1} \approx \mathbf{1} - \mathbf{V}\cdot\mathbf{M}$, it is found that:

$$\mathbf{M}' \approx \mathbf{M} + \left[\frac{1}{c_0^2}\mathbf{V} - \mathbf{M}\cdot\mathbf{V}\cdot\mathbf{M}\right] = \begin{pmatrix} \boldsymbol{\varepsilon}(\mathbf{r}') & \frac{1}{c_0^2}\mathbf{v}\times\mathbf{1} - \boldsymbol{\varepsilon}\cdot\left[\mathbf{v}\times\boldsymbol{\mu}\right] \\ -\frac{1}{c_0^2}\mathbf{v}\times\mathbf{1} + \boldsymbol{\mu}\cdot\left[\mathbf{v}\times\boldsymbol{\varepsilon}\right] & \boldsymbol{\mu}(\mathbf{r}') \end{pmatrix}. \tag{S8}$$

Thus, in the non-relativistic limit one obtains:

$$\boldsymbol{\varepsilon}' \approx \boldsymbol{\varepsilon}(\mathbf{r}'), \qquad \boldsymbol{\mu}' \approx \boldsymbol{\mu}(\mathbf{r}'), \qquad \boldsymbol{\xi}' = \boldsymbol{\zeta}'^{T} \approx \frac{1}{c_0^2}\mathbf{v}\times\mathbf{1} - \boldsymbol{\varepsilon}(\mathbf{r}')\cdot\left[\mathbf{v}\times\boldsymbol{\mu}(\mathbf{r}')\right]. \tag{S9}$$

As seen, the permittivity and permeability are unaffected by the coordinate transformation. Using Eq. (S7) and $\mathbf{v} = v_x\hat{\mathbf{x}} + v_y\hat{\mathbf{y}}$, one finds that the relevant entries of the magneto-electric tensor can be written explicitly as:

$$\xi'_{zx} \approx -\varepsilon_{zz}\mu_{xy}v_x + \left(\varepsilon_{zz}\mu_{xx} - \frac{1}{c_0^2}\right)v_y, \tag{S10a}$$

$$\xi'_{zy} \approx -\left(\varepsilon_{zz}\mu_{yy} - \frac{1}{c_0^2}\right)v_x + \varepsilon_{zz}\mu_{xy}v_y. \tag{S10b}$$

Note that the magneto-electric parameters depend explicitly on $c_0$ and thus are gauge dependent, i.e., transformation dependent.

*II. Exact result for* $\mathbf{v} = v_y \hat{\mathbf{y}}$

For completeness, we also show the exact formulas for $\varepsilon'_{zz}$, $\mu'_{xx}$, $\mu'_{yy}$, $\mu'_{xy}$, $\xi'_{zx}$, $\xi'_{zy}$ for a modulation speed along the *y*-direction $\mathbf{v} = v\hat{\mathbf{y}}$:

$$\varepsilon'_{zz} = \frac{\varepsilon_{zz}\left(1 - v^2 / c_0^2\right)}{1 - \varepsilon_{zz}\mu_{xx}v^2}, \tag{S11a}$$

$$\mu'_{xx} = \frac{\mu_{xx}\left(1 - v^2 / c_0^2\right)}{1 - \varepsilon_{zz}\mu_{xx}v^2}, \quad \mu'_{xy} = \sqrt{1 - \frac{v^2}{c_0^2}} \frac{\mu_{xy}}{1 - \varepsilon_{zz}\mu_{xx}v^2}, \quad \mu'_{yy} = \frac{\mu_{yy} - \varepsilon_{zz}\left(\mu_{xx}\mu_{yy} - \mu_{xy}^2\right)v^2}{1 - \varepsilon_{zz}\mu_{xx}v^2} \tag{S11b}$$

$$\xi'_{zx} = \left(\varepsilon_{zz}\mu_{xx} - \frac{1}{c_0^2}\right)\frac{v}{1 - \varepsilon_{zz}\mu_{xx}v^2}, \qquad \xi'_{zy} = \sqrt{1 - \frac{v^2}{c_0^2}}\varepsilon_{zz}\mu_{xy}\frac{v}{1 - \varepsilon_{zz}\mu_{xx}v^2}. \tag{S11c}$$

It is underlined that the above formulas apply both to a Lorentz transformation ($c_0 = c$) and to a Galilean transformation ($c_0 = \infty$).

## B. Gauge transformations

As discussed in the Sect. A, there is an arbitrariness in the choice of the coordinate transformation. In fact, the parameter $c_0$ can be taken arbitrarily in the range $v < c_0 \leq \infty$. All the transformations predict the same electrodynamics in the laboratory frame when the corresponding inverse transformation is applied to the fields and to the spacetime coordinates.

Evidently, the state vector in the co-moving frame depends on the parameter $c_0$, i.e., $\mathbf{\Psi}' = \mathbf{\Psi}'_{c_0}$ (explicit formulas for the co-moving frame fields can be found in Ref. [27]). In the following, we prove that in the non-relativistic limit different state vectors $\mathbf{\Psi}'_{c_0}$ are linked by a gauge transformation, such that

$$\mathbf{\Psi}'_{c_0} \equiv \mathbf{\Psi}'_{\infty} e^{-\frac{\omega'}{c_0^2} i \left(v_x x' + v_y y'\right)}, \tag{S12}$$

where $\mathbf{v} = v_x \hat{\mathbf{x}} + v_y \hat{\mathbf{y}}$ is the modulation speed and $\mathbf{\Psi}'_{\infty}$ is the state vector for the Galilean case ($c_0 = \infty$).

To begin with, we note that the eigenstates $\mathbf{\Psi}'_{\infty}$ of the operator obtained with a Galilean transformation satisfy:

$$\begin{pmatrix} 0 & -i\partial_{y'} & +i\partial_{x'} \\ -i\partial_{y'} & 0 & 0 \\ +i\partial_{x'} & 0 & 0 \end{pmatrix} \cdot \mathbf{\Psi}'_{\infty} = \omega' \begin{pmatrix} \varepsilon_{zz}(\mathbf{r}') & \xi'_{zx,\mathrm{G}}(\mathbf{r}') & \xi'_{zy,\mathrm{G}}(\mathbf{r}') \\ \xi'_{zx,\mathrm{G}}(\mathbf{r}') & \mu_{xx}(\mathbf{r}') & \mu_{xy}(\mathbf{r}') \\ \xi'_{zy,\mathrm{G}}(\mathbf{r}') & \mu_{xy}(\mathbf{r}') & \mu_{yy}(\mathbf{r}') \end{pmatrix} \cdot \mathbf{\Psi}'_{\infty}, \tag{S13}$$

with $\omega'$ the oscillation eigenfrequency in the Galilean co-moving frame. We combined Eq. (2) of the main text with Eq. (S10) and $\boldsymbol{\varepsilon}' \approx \boldsymbol{\varepsilon}(\mathbf{r}')$, $\boldsymbol{\mu}' \approx \boldsymbol{\mu}(\mathbf{r}')$. In the above, $\xi'_{zx,\mathrm{G}} \approx -\varepsilon_{zz}\mu_{xy}v_x + \varepsilon_{zz}\mu_{xx}v_y$ and $\xi'_{zy,\mathrm{G}} \approx -\varepsilon_{zz}\mu_{yy}v_x + \varepsilon_{zz}\mu_{xy}v_y$ are the magneto-electric parameters evaluated with $c_0 = \infty$. From here, it is simple to check that $\mathbf{\Psi}'_{c_0}$ defined as in Eq. (S12) satisfies:

$$\begin{pmatrix} 0 & -i\partial_{y'} & +i\partial_{x'} \\ -i\partial_{y'} & 0 & 0 \\ +i\partial_{x'} & 0 & 0 \end{pmatrix} \cdot \mathbf{\Psi}'_{c_0} = \omega' \begin{pmatrix} \varepsilon_{zz}(\mathbf{r}') & \xi'_{zx,\mathrm{G}}(\mathbf{r}') - \frac{v_y}{c_0^2} & \xi'_{zy,\mathrm{G}}(\mathbf{r}') + \frac{v_x}{c_0^2} \\ \xi'_{zx,\mathrm{G}}(\mathbf{r}') - \frac{v_y}{c_0^2} & \mu_{xx}(\mathbf{r}') & \mu_{xy}(\mathbf{r}') \\ \xi'_{zy,\mathrm{G}}(\mathbf{r}') + \frac{v_x}{c_0^2} & \mu_{xy}(\mathbf{r}') & \mu_{yy}(\mathbf{r}') \end{pmatrix} \cdot \mathbf{\Psi}'_{c_0}. \tag{S14}$$

But since $\xi'_{zx} = \xi'_{zx,\mathrm{G}} - \frac{v_y}{c_0^2}$ and $\xi'_{zy} = \xi'_{zy,\mathrm{G}} + \frac{v_x}{c_0^2}$, it follows that the eigenstates of the operator constructed with a generalized Lorentz $c_0$-transformation are linked to the eigenstates obtained with a Galilean transformation as in Eq. (S12). Thus, $\mathbf{\Psi}'$ can be understood as a gauge field. The gauge degree freedom is associated with the arbitrariness of the parameter $c_0$. It is underlined that even though the eigenstates in the co-moving frame depend on the gauge (i.e., depend on $c_0$), the corresponding fields in the laboratory frame are by construction gauge independent.

## *C. Synthetic magnetic potential*

### *I. Non-relativistic limit*

As shown in Refs. [26, 30, 31] based on an analogy with the Schrödinger equation, the bianisotropic coupling determines a synthetic magnetic potential. For TE-polarized waves, it is given by $\mathbf{A} = \left[ c\xi'_{zx}\hat{\mathbf{x}} + c\xi'_{zy}\hat{\mathbf{y}} \right] \times \hat{\mathbf{z}}$ (see Appendix B of Ref. [26]).

For simplicity, in the following discussion we focus on the nonrelativistic limit. In such a case, $\xi'_{zx}$, $\xi'_{zy}$ can be evaluated using Eq. (S10). Hence, the magnetic potential reduces to:

$$\frac{\mathbf{A}}{c} \approx \left[ -\left( \varepsilon_{zz}\mu_{yy} - \frac{1}{c_0^2} \right) v_x + \varepsilon_{zz}\mu_{xy} v_y \right] \hat{\mathbf{x}} + \left[ \varepsilon_{zz}\mu_{xy} v_x - \left( \varepsilon_{zz}\mu_{xx} - \frac{1}{c_0^2} \right) v_y \right] \hat{\mathbf{y}} . \qquad \text{(S15)}$$

In the particular case of an isotropic material, $\mu \equiv \mu_{xx} = \mu_{yy}$, $\mu_{xy} = 0$, $\varepsilon \equiv \varepsilon_{zz}$ the vector potential can be written as $\mathbf{A} \approx -\left( n^2 - c^2 / c_0^2 \right) \mathbf{v} / c$ with $n = c\sqrt{\varepsilon\mu}$ the refractive index of the material. Thus, in the isotropic case the direction of the vector potential is locked to the direction of the modulation speed. Note that for a Lorentz transformation ($c_0 = c$) the magnetic potential reduces to $\mathbf{A} \approx -\left(n^2-1\right)\mathbf{v}/c$ which is the result discussed in the main text.

*II. Perturbation of an isorefractive crystal*

Consider now an isotropic spacetime crystal ( $\mu \equiv \mu_{xx} = \mu_{yy}$, $\mu_{xy} = 0$, $\varepsilon \equiv \varepsilon_{zz}$ ) formed by isorefractive materials so that $n = c\sqrt{\varepsilon\mu} = const.$. Then, there is a gauge for which the magneto-electric coefficients $\xi'_{zx}$, $\xi'_{zy}$ vanish exactly [27]. Specifically, the relevant gauge is associated with the generalized Lorentz transformation with $c_0 = c/n$ [see Eq. (S11c)]. This result is exact, i.e., it does not require a linear $v$ approximation. Isorefractive media with $n = const.$ are "fixed points" of the generalized $c_0 = c/n$ Lorentz transformation, in the sense that they stay invariant under the coordinate transformation. In particular, it was recently shown that spacetime crystals formed by materials with $n=1$ behave exactly as moving photonic crystals [27]. Such crystals are designated by Minkowskian spacetime crystals.

It is evident that any isorefractive spacetime crystal must be topologically trivial. In fact, the crystal response in a suitable co-moving frame ($c_0 = c/n$) is reciprocal ( $\xi'_{zx} = 0 = \xi'_{zy}$ ). Furthermore, the synthetic magnetic potential vanishes in the generalized Lorentz $c_0 = c/n$ co-moving frame. For a generic coordinate transformation, $\mathbf{A} \approx \left(n^2 - c^2/c_0^2\right)\mathbf{v}/c = const.$, which corresponds to a trivial synthetic magnetic field ( $\nabla \times \mathbf{A} = 0$ ), in agreement with the trivial topology.

It is interesting to study how the magnetic potential changes if one perturbs slightly an isorefractive spacetime crystal. To investigate this, we assume that the magnetic response may be slightly anisotropic so that the in-plane permeability is of the form:

$$\boldsymbol{\mu}_{\text{in-plane}}(\mathbf{r},t)=\mu\left[\left(1+\delta_\mu\right)\hat{\mathbf{e}}_1\otimes\hat{\mathbf{e}}_1+\left(1-\delta_\mu\right)\hat{\mathbf{e}}_2\otimes\hat{\mathbf{e}}_2\right]. \tag{S16}$$

Here, $\hat{\mathbf{e}}_1$, $\hat{\mathbf{e}}_2$ are the in-plane orthogonal permeability main axes, defined so that $\hat{\mathbf{e}}_1\times\hat{\mathbf{e}}_2=\hat{\mathbf{z}}$. The in-plane permeability eigenvalues are $\mu\left(1\pm\delta_\mu\right)$. All the parameters ($\hat{\mathbf{e}}_1$, $\hat{\mathbf{e}}_2$, $\mu$, $\delta_\mu$) are functions of $\mathbf{r}-\mathbf{v}t$ in the laboratory frame coordinates. Furthermore, it is supposed that $\varepsilon_{zz}\mu=const.\equiv n^2/c^2$ so that when $\delta_\mu=0$ the crystal is isorefractive. Thus, $\delta_\mu$ represents the detuning with respect to the isorefractive condition.

Let us denote $\hat{\mathbf{e}}_1=\cos\left(\varphi_\mu\right)\hat{\mathbf{x}}+\sin\left(\varphi_\mu\right)\hat{\mathbf{y}}$, $\hat{\mathbf{e}}_2=-\sin\left(\varphi_\mu\right)\hat{\mathbf{x}}+\cos\left(\varphi_\mu\right)\hat{\mathbf{y}}$, so that $\varphi_\mu$ represents the angle of the main axis $\hat{\mathbf{e}}_1$ with respect to the *x*-direction (see Fig. 2a). Then, we have:

$$\mu_{xx}=\mu\left(1+\delta_\mu\right)\cos^2\left(\varphi_\mu\right)+\mu\left(1-\delta_\mu\right)\sin^2\left(\varphi_\mu\right)=\mu\left[1+\delta_\mu\cos\left(2\varphi_\mu\right)\right], \tag{S17a}$$

$$\mu_{yy}=\mu\left(1+\delta_\mu\right)\sin^2\left(\varphi_\mu\right)+\mu\left(1-\delta_\mu\right)\cos^2\left(\varphi_\mu\right)=\mu\left[1-\delta_\mu\cos\left(2\varphi_\mu\right)\right]. \tag{S17b}$$

$$\mu_{xy}=2\mu\delta_\mu\cos\left(\varphi_\mu\right)\sin\left(\varphi_\mu\right)=\mu\delta_\mu\sin\left(2\varphi_\mu\right). \tag{S17c}$$

Let us also write $\mathbf{v}=v\left[\cos\left(\varphi_v\right)\hat{\mathbf{x}}+\sin\left(\varphi_v\right)\hat{\mathbf{y}}\right]$ so that $\varphi_v$ represents the angle of the modulation speed $\mathbf{v}$ with respect to the $x$-direction (see Fig. 2a). Substituting these formulas in Eq. (S15), one finds that the synthetic magnetic potential is:

$$\mathbf{A} \approx -\frac{\mathbf{v}}{c}\left(n^2 - \frac{c^2}{c_0^2}\right) + n^2 \frac{v}{c}\delta_\mu \hat{\mathbf{e}}_{\mathrm{A}} . \tag{S18}$$

The first term on the right-hand side is the constant synthetic potential for the isorefractive crystal, which has no impact on the topology. The interesting piece is the second term. Its direction is ruled by the relative orientation of the main axes of the permeability with respect to the modulation speed: $\hat{\mathbf{e}}_{\mathrm{A}} \equiv \cos\left(2\varphi_\mu - \varphi_v\right)\hat{\mathbf{x}} + \sin\left(2\varphi_\mu - \varphi_v\right)\hat{\mathbf{y}}$. In particular, if we pick the gauge $c_0 = c/n$ the magnetic potential becomes $\mathbf{A} \approx n^2 \frac{v}{c}\delta_\mu \hat{\mathbf{e}}_{\mathrm{A}}$. In the main text, we construct a spacetime crystal that corresponds to the perturbation of a Minkowskian crystal with $n = 1$. In this case, the vector potential reduces to $\mathbf{A} \approx \frac{v}{c}\delta_\mu \hat{\mathbf{e}}_{\mathrm{A}}$ in the Lorentz co-moving frame.

### *D. Band structure of the spacetime crystal*

The electromagnetic modes in the co-moving frame coordinates are regular Bloch waves. This is so because the material matrix $\mathbf{M}'_{\mathrm{TE}}$ is periodic in the primed coordinates. From Eq. (2), the TE Bloch waves with envelope $\mathbf{\Psi}_{\mathrm{p}}$ are solutions of the generalized eigenvalue problem:

$$\underbrace{\begin{pmatrix} 0 & -i\partial_{y'} + k'_y & +i\partial_{x'} - k'_x \\ -i\partial_{y'} + k'_y & 0 & 0 \\ +i\partial_{x'} - k'_x & 0 & 0 \end{pmatrix}}_{\mathbf{L}_{\mathbf{k}'}} \cdot \mathbf{\Psi}_{\mathrm{p}} = \omega' \underbrace{\begin{pmatrix} \varepsilon'_{zz} & \xi'_{zx} & \xi'_{zy} \\ \xi'_{zx} & \mu'_{xx} & \mu'_{xy} \\ \xi'_{zy} & \mu'_{xy} & \mu'_{yy} \end{pmatrix}}_{\mathbf{M}'_{\mathrm{TE}}} \cdot \mathbf{\Psi}_{\mathrm{p}} . \tag{S19}$$

Here, $\mathbf{k}' = k'_x \hat{\mathbf{x}} + k'_y \hat{\mathbf{y}}$ is the wave vector of the Bloch wave and $\omega'$ is the frequency.

The band structure in the co-moving frame coordinates can be found using a plane wave expansion:

$$\mathbf{\Psi}_{\mathrm{p}}\left(\mathbf{r}'\right) = \sum_{\mathbf{q}\in\mathbb{Z}^2} \boldsymbol{\psi}_{\mathbf{q}} e^{i\mathbf{G}^0_{\mathbf{q}}\cdot\mathbf{r}'} \quad \boldsymbol{\psi}_{\mathbf{q}} = \begin{pmatrix} \psi_{1,\mathbf{q}} \\ \psi_{1,\mathbf{q}} \\ \psi_{1,\mathbf{q}} \end{pmatrix}. \tag{S20}$$

where $\mathbf{G}^0_{\mathbf{q}} = q_1\mathbf{b}'_1 + q_2\mathbf{b}'_2$ is a generic vector of the reciprocal lattice. The primitive vectors of the reciprocal and direct lattices are related in the usual manner $\mathbf{b}'_i \cdot \mathbf{a}'_j = 2\pi\delta_{ij}$. Furthermore, since the material matrix is a periodic function, it can also be expanded into plane waves as:

$$\mathbf{M}'_{\mathrm{TE}}\left(\mathbf{r}'\right) = \sum_{\mathbf{q}\in\mathbb{Z}^2} \mathbf{M}'_{\mathbf{q}} e^{i\mathbf{G}^0_{\mathbf{q}}\cdot\mathbf{r}'}, \qquad \text{with} \qquad \mathbf{M}'_{\mathbf{q}} = \begin{pmatrix} \varepsilon'_{zz,\mathbf{q}} & \xi'_{zx,\mathbf{q}} & \xi'_{zy,\mathbf{q}} \\ \xi'_{zx,\mathbf{q}} & \mu'_{xx,\mathbf{q}} & \mu'_{xy,\mathbf{q}} \\ \xi'_{zy,\mathbf{q}} & \mu'_{xy,\mathbf{q}} & \mu'_{yy,\mathbf{q}} \end{pmatrix}. \tag{S21}$$

When the material inclusions have a circular cross-section (as in the main text), the Fourier coefficients $\mathbf{M}_{\mathbf{q}}$ can be expressed in terms of cylindrical Bessel functions, similar to Refs. [26, 31, 38] (not shown). Substituting the previous formulas into Eq. (S19), one obtains a generalized matrix eigenvalue problem:

$$\begin{pmatrix} [0] & \left[\mathbf{G}_{\mathbf{p}}\cdot\hat{\mathbf{y}}\delta_{\mathbf{p}-\mathbf{q}}\right] & \left[-\mathbf{G}_{\mathbf{p}}\cdot\hat{\mathbf{x}}\delta_{\mathbf{p}-\mathbf{q}}\right] \\ \left[\mathbf{G}_{\mathbf{p}}\cdot\hat{\mathbf{y}}\delta_{\mathbf{p}-\mathbf{q}}\right] & [0] & [0] \\ \left[-\mathbf{G}_{\mathbf{p}}\cdot\hat{\mathbf{x}}\delta_{\mathbf{p}-\mathbf{q}}\right] & [0] & [0] \end{pmatrix} \cdot \begin{pmatrix} \left(\psi_{1,\mathbf{q}}\right) \\ \left(\psi_{2,\mathbf{q}}\right) \\ \left(\psi_{3,\mathbf{q}}\right) \end{pmatrix} = \omega' \begin{pmatrix} \left[\varepsilon'_{zz,\mathbf{p}-\mathbf{q}}\right] & \left[\xi'_{zx,\mathbf{p}-\mathbf{q}}\right] & \left[\xi'_{zy,\mathbf{p}-\mathbf{q}}\right] \\ \left[\xi'_{zx,\mathbf{p}-\mathbf{q}}\right] & \left[\mu'_{xx,\mathbf{p}-\mathbf{q}}\right] & \left[\mu'_{xy,\mathbf{p}-\mathbf{q}}\right] \\ \left[\xi'_{zy,\mathbf{p}-\mathbf{q}}\right] & \left[\mu'_{xy,\mathbf{p}-\mathbf{q}}\right] & \left[\mu'_{yy,\mathbf{p}-\mathbf{q}}\right] \end{pmatrix} \cdot \begin{pmatrix} \left(\psi_{1,\mathbf{q}}\right) \\ \left(\psi_{2,\mathbf{q}}\right) \\ \left(\psi_{3,\mathbf{q}}\right) \end{pmatrix} \tag{S22}$$

with $\mathbf{G}_{\mathbf{q}} = \mathbf{k}' + \mathbf{G}^0_{\mathbf{q}}$ and $\mathbf{p},\mathbf{q} \in \mathbb{Z}^2$. The sub-blocks $\left[A_{\mathbf{p},\mathbf{q}}\right]$ and $\left(F_{\mathbf{q}}\right)$ represent square matrices and vectors, respectively, with infinite dimension. In practice, the plane wave expansion is truncated so that all the matrices have a finite rank. It can be shown that the material matrix is positive definite in the subluminal regime, i.e., when the modulation speed is smaller than the speed of light in the considered materials (as in all the examples of this work). Thus, the generalized eigenvalue problem can be reduced to a standard matrix eigensystem using a Cholesky decomposition of the material matrix.

### *E. Calculation of the gap Chern number*

The topological charge of the spacetime crystal is calculated in the co-moving frame coordinates. In the co-moving frame, the spectral problem reduces to a standard eigensystem as shown in Eq. (S19). The solutions are Bloch waves defined in the Brillouin zone ( B.Z. ) of the reciprocal space. The Berry curvature can be calculated from the Green's function of the system [26, 31, 38] defined as:

$$\mathbf{G}_{\mathbf{k}'}(\omega') = i\left(\mathbf{L}_{\mathbf{k}'} - \omega'\mathbf{M}'_{\mathrm{TE}}\right)^{-1}, \tag{S23}$$

with the operators $\mathbf{L}_{\mathbf{k}'}$, $\mathbf{M}'_{\mathrm{TE}}$ determined by Eq. (S19). At a given point $\mathbf{k}' \in \mathrm{B.Z.}$, the Berry curvature is found by integrating the Green's function in the complex frequency plane [26, 31, 38]:

$$\mathcal{F}_{\mathbf{k}'} = \frac{i}{(2\pi)^2} \int_{\omega'_{\mathrm{gap}}-i\infty}^{\omega'_{\mathrm{gap}}+i\infty} d\omega' \; \mathrm{Tr}\left\{ \frac{\partial \mathbf{L}_{\mathbf{k}'}}{\partial k'_x} \cdot \mathbf{G}_{\mathbf{k}'}(\omega') \cdot \frac{\partial \mathbf{L}_{\mathbf{k}'}}{\partial k'_y} \cdot \mathbf{G}_{\mathbf{k}'}(\omega') \cdot \mathbf{M}'_{\mathrm{TE}} \cdot \mathbf{G}_{\mathbf{k}'}(\omega') \right\}. \tag{S24}$$

The integration path is the line $\mathrm{Re}\{\omega'\} = \omega'_{\mathrm{gap}}$ parallel to the imaginary frequency axis, with $\omega'_{\mathrm{gap}}$ some frequency in the relevant band gap. The gap Chern number is found by integrating the Berry curvature over the Brillouin zone:

$$\mathcal{C}_{\mathrm{gap}} = \iint_{\mathrm{B.Z.}} \mathcal{F}_{\mathbf{k}'} d^2\mathbf{k}' . \tag{S25}$$

In the numerical calculations, all the operators are represented by the matrices obtained with the plane wave expansion.

The topological charge is independent of the coordinate transformation (independent of $c_0$) because the eigenfunctions associated with different generalized Lorentz transformations are linked as $\mathbf{\Psi}' \to \mathbf{\Psi}' e^{i\mathbf{k}_\Delta \cdot \mathbf{r}'}$ with $\mathbf{k}_\Delta$ some vector independent of $\mathbf{r}'$ (see the supplementary note B).

Figure S1a shows a density plot of the Berry curvature [Eq. (S24)] for the band gap highlighted in Fig. 3b of the main text. We integrate Eq. (S24) using $\mathcal{F}_{\mathbf{k}'} = \frac{-1}{(2\pi)^2} \int_{-\infty}^{+\infty} d\xi \; \mathrm{Tr}\{...\}_{\omega' = \omega'_{\mathrm{gap}} + i\xi}$. The integration is performed numerically using the trapezoidal rule. The integrand is discretized into $N_\xi$ sampling points and the integration range is truncated to an interval of the type $-\xi_{\max} < \xi < \xi_{\max}$. As seen, the Berry curvature is strongly concentrated near two points of the reciprocal space, which coincide approximately with the Dirac points. Note that in Fig. S1 we parameterize the Brillouin zone using $\mathbf{k}' = \beta_1 \mathbf{b}'_1 + \beta_2 \mathbf{b}'_2$ with $|\beta_i| < 1/2$.

Furthermore, the gap Chern number is numerically evaluated using $\mathcal{C}_{\mathrm{gap}} = \int_{-1/2}^{1/2} d\beta_1 \int_{-1/2}^{1/2} d\beta_2 \; \mathcal{F}_{\mathbf{k}' = \beta_1 \mathbf{b}'_1 + \beta_2 \mathbf{b}'_2} \left| \mathbf{b}'_1 \times \mathbf{b}'_2 \right|$ with the Brillouin Zone discretized into $N_x \times N_y$ points. Figure S1b shows the convergence of the numerically calculated gap Chern number as a function of $N_x = N_y$ for $\xi_{\max} = 0.1 c/a$ and $N_\xi = 100$, confirming that for sufficiently large $N_x = N_y$ the gap Chern number approaches an integer. We performed similar convergence tests in $\xi_{\max}$ and $N_\xi$.

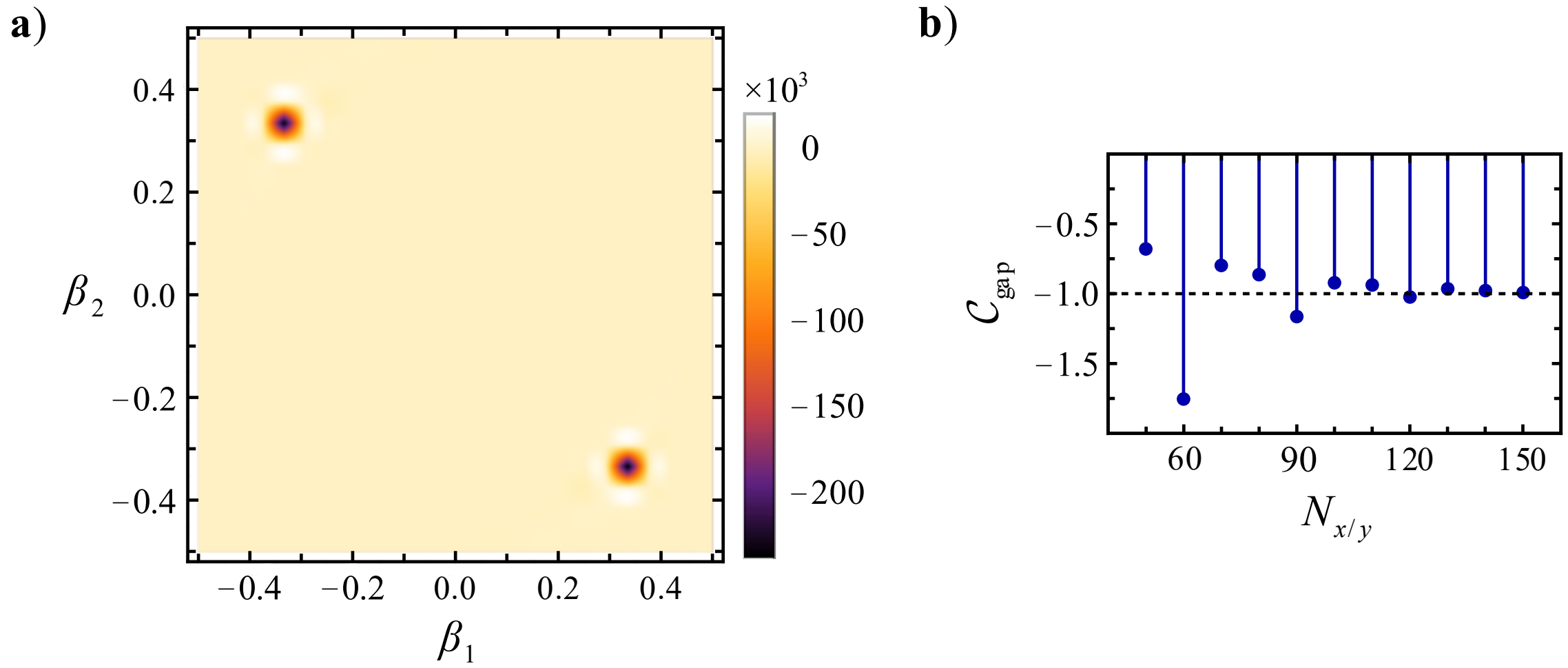

**Fig. S1 a)** Normalized Berry curvature $|\mathbf{b}_1' \times \mathbf{b}_2'| \mathcal{F}_{\mathbf{k}'}$ with $\mathbf{k}' = \beta_1 \mathbf{b}_1' + \beta_2 \mathbf{b}_2'$ for the band gap shaded in blue in Fig. 3b of the main text. **b)** Gap Chern number convergence analysis as a function of the number of points $N_{x/y}$ used to discretize each direction of the Brillouin Zone.

## *F. Symmetry constraints on the emergence of unidirectional edge modes*

In this supplementary note, we discuss how some discrete symmetries constraint the formation of topological states.

It is well known that a system that possesses either a time-reversal symmetry ($\mathcal{T}$), or a mirror symmetry (e.g., a mirror symmetry with respect to the $x$-direction $\mathcal{P}_x : (x, y) \to (-x, y)$) or a parity-time symmetry ($\mathcal{P} \cdot \mathcal{T}$, with $\mathcal{P} : (x, y) \to (-x, -y)$ corresponding to a two-fold rotation) is necessarily topologically trivial. Formally, this happens because the Berry curvature has an odd symmetry in the Brillouin Zone and, consequently, the gap Chern number is trivial. For example, for a $\mathcal{P}_x$-symmetric system, the Berry curvature satisfies $\mathcal{F}_{(k_x,k_y)} = -\mathcal{F}_{(-k_x,k_y)}$ and hence its integral over the Brillouin zone vanishes.

It is interesting to justify these properties in a more geometrical way. To this end, in the following we discuss how a given symmetry constraints the net number of unidirectional edge modes at an interface with a trivial topological insulator (e.g., a perfectly electric conducting, PEC, wall). This approach based on the bulk-edge correspondence provides a simple and intuitive method to grasp how different symmetries may constraint the design of non-trivial topologies.

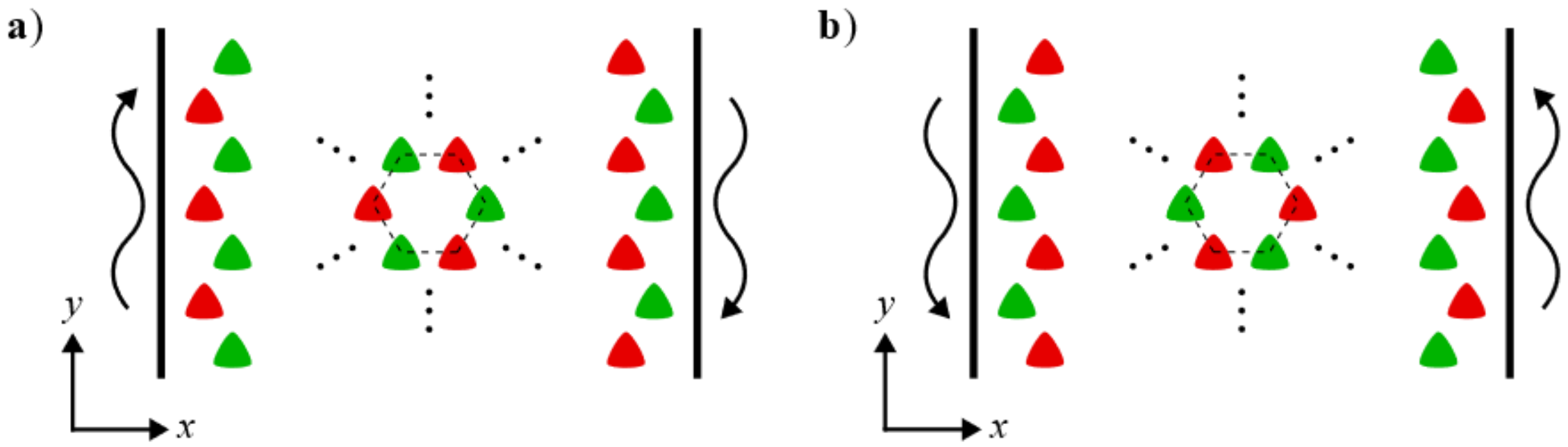

**Fig. S2 a)** Illustration of a photonic crystal with a honeycomb lattice structure enclosed in a cavity with PEC walls (represented in black). The black arrows represent the propagation of an edge mode in the clockwise direction around the cavity walls. The unit cell of the unbounded crystal is formed by two inequivalent elements represented in green and red colors. **b)** Photonic crystal in a) after the action of the mirror operator $\mathcal{P}_x$: the red elements switch position with the green elements and the edge mode propagates in the counterclockwise direction.

Let us take the mirror symmetry $\mathcal{P}_x$ as an example. Figure S2a represents a photonic crystal with a honeycomb structure enclosed in a large metallic cavity with PEC walls. From the bulk-edge correspondence, the net number of unidirectional states circulating in the counterclockwise direction around the cavity is identical to $-\mathcal{C}_{\text{gap}}$, with $\mathcal{C}_{\text{gap}}$ the topological charge of the band gap [18]. The unit cell of the unbounded crystal is formed by two inequivalent elements depicted in green and red in Fig. S2.

Under the $\mathcal{P}_x$ mirror operation, the considered photonic crystal is transformed into a new crystal where the positions of the red and green elements are interchanged (Fig. S2b). Furthermore, under the $\mathcal{P}_x$ operation, all the edge modes propagating in the counterclockwise (clockwise) direction are transformed into modes circulating in the clockwise (counterclockwise) direction. This property implies that the gap Chern number of the $\mathcal{P}_x$-transformed system has the opposite sign as the gap Chern number of the original system. In particular, when the system is $\mathcal{P}_x$-invariant, the gap Chern number must be exactly zero, and thereby such systems cannot support unidirectional edge modes. The same analysis can be done for other symmetries.

It is interesting to analyze in further detail how the mirror symmetry constraints the formation of nontrivial topologies in spacetime crystals. To this end, we consider again a crystal with a honeycomb lattice but now subject to a spacetime modulation with the modulation velocity directed along the *y*-direction (Fig. S3). As before, the crystal is enclosed in a cavity with opaque walls. If the red and green elements are different, the crystal breaks simultaneously $\mathcal{T}$ and $\mathcal{P}\cdot\mathcal{T}$ symmetries.

Let us focus on the mirror symmetry $\mathcal{P}_x : (x, y) \rightarrow (-x, y)$, so that the modulation velocity ( $\mathbf{v} = v\hat{\mathbf{y}}$ ) is parallel to the mirror. Under the action of $\mathcal{P}_x$, the system of Fig. S3a is transformed into the system of Fig. S3b. As seen, the $\mathcal{P}_x$ mirror operator preserves the modulation speed but flips the direction of circulation of all the edge modes. In particular, it follows that to have a nontrivial topology the unit cell cannot have $\mathcal{P}_x$-mirror symmetry. The mirror symmetry can be broken with scatterers with a triangular shape, as shown in Fig. S3. This example clearly shows that a nontrivial topology requires a unit cell without any mirror-plane parallel to the modulation velocity. In contrast, mirrors perpendicular to the direction of the modulation speed do not constraint the topological charge of the spacetime crystal.

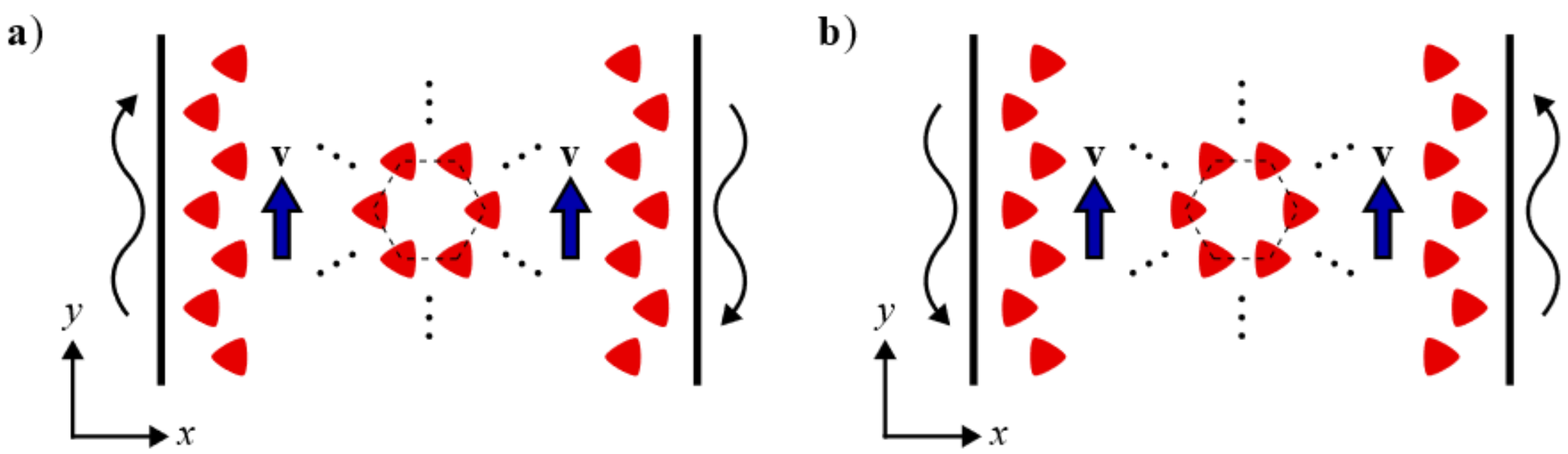

**Fig. S3 a)** Similar to Fig. S2a but for a spacetime crystal with a traveling-wave modulation with velocity $\mathbf{v} = v\hat{\mathbf{y}}$. **b)** Photonic crystal in a) after the action of the mirror symmetry operator $\mathcal{P}_x$: the modulation velocity is preserved, but the direction of propagation of the edge modes is flipped. To generate non-trivial topologies, the unit cell must have a broken $\mathcal{P}_x$- symmetry (as in the figure).

## *G. Spacetime crystals with a square lattice*

In this supplementary note, we study the band structure and topology of a spacetime crystal with dielectric elements organized in a square lattice (Fig. S4a). The modulation velocity is along the $y$-direction.

From the supplementary note F, a nontrivial topology is possible only if the unit cell does not have any mirror-plane parallel to $\mathbf{v} = v\hat{\mathbf{y}}$. To ensure this, the dielectric scatterers have a diamond-shape cross-section with a lateral “cut” (red structures in Fig. S4a). In Ref. [34], it was shown that a square lattice may exhibit two pairs of Dirac cones, one protected by $\mathcal{P}_x$ symmetry and the other one by $\mathcal{P}_y$ symmetry. Since the geometry of our inclusions breaks $\mathcal{P}_x$ symmetry, the photonic band structure in the static case ($v = 0$) presents only one pair of Dirac cones over the high-symmetry lines $Y - M$ protected by $\mathcal{P}_y$ symmetry (Fig. S4b). When the spacetime modulation is switched on ($v \neq 0$), the Dirac degeneracies are lifted and a complete band gap is formed (Fig. S4c). However, it turns out that the band gap is topologically trivial ($\mathcal{C}_{\text{gap}} = 0$), similar to the example in Fig. 1 of the main text. Figure S4d shows that the topological charge density is concentrated at the two Dirac points on the high-symmetry lines $X - M$ ($\beta_1 = \pm 0.5$). The topological charge density has an opposite sign at each Dirac point, hence the trivial Chern topology. Thus, a broken $\mathcal{P}_x$-mirror symmetry and a broken 2-fold rotation symmetry do not guarantee the emergence of nontrivial topological phases.

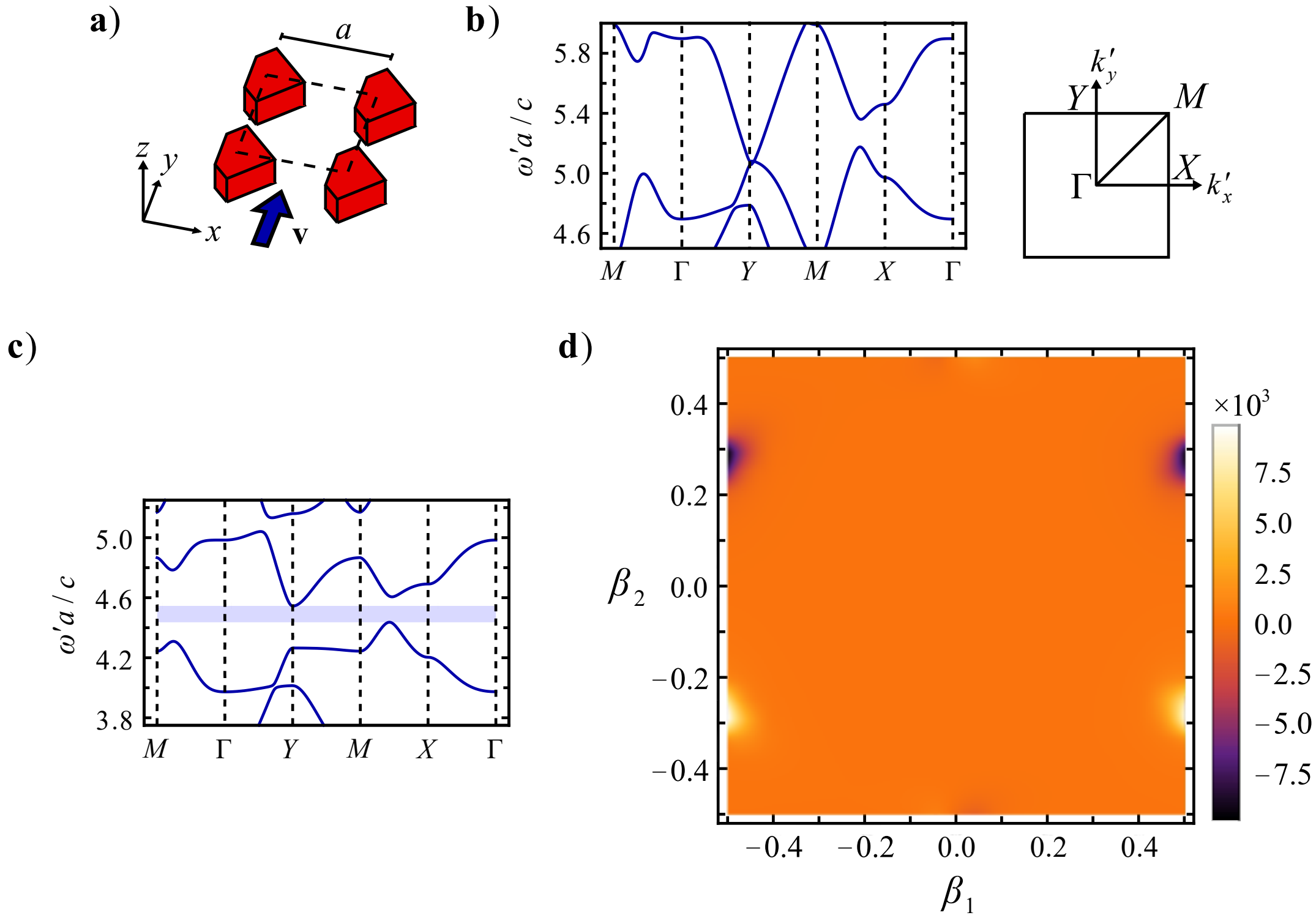

**Fig. S4 a)** Unit cell of a spacetime crystal with the dielectric scatterers organized in a square lattice in the Lorentz co-moving frame. The modulation velocity is $\mathbf{v} = v\hat{\mathbf{y}}$. The dielectric inclusion ($\varepsilon = 8\varepsilon_0$, $\mu = \mu_0$) has a diamond-like geometry with a lateral cut that breaks the $\mathcal{P}_x$ symmetry. **b)** (Left) Band structure in the Lorentz co-moving frame for $v = 0$. (Right) Brillouin Zone in the dual space of the Lorentz co-moving frame. **c)** Band structure for $v = 0.2c$. The band gap corresponding to the horizontal blue strip is characterized by $\mathcal{C}_{\text{gap}} = 0$. **d)** Normalized Berry curvature $|\mathbf{b}'_1 \times \mathbf{b}'_2| \mathcal{F}_{\mathbf{k}'}$ with $\mathbf{k}' = \beta_1 \mathbf{b}'_1 + \beta_2 \mathbf{b}'_2$ for the band gap shaded in blue in c).

## *H. Other implementations of the "anisotropic ring"*

In this supplementary note, we discuss different approximations of the anisotropic ring introduced in the main text (see Fig. 2bi). Similar to the main text, the continuous ring is approximated by an array of discrete scatterers.

We start with the simplest and crudest approximation of the ring, using only $N=2$ scatterers (bluish elements in Fig. S5a). Similar to the main text, the "ring" perturbs slightly a Minkowskian isorefractive crystal with $n=1$ corresponding to the red elements embedded in vacuum. The modulation velocity is along $y$ ($\mathbf{v}=v\hat{\mathbf{y}}$).

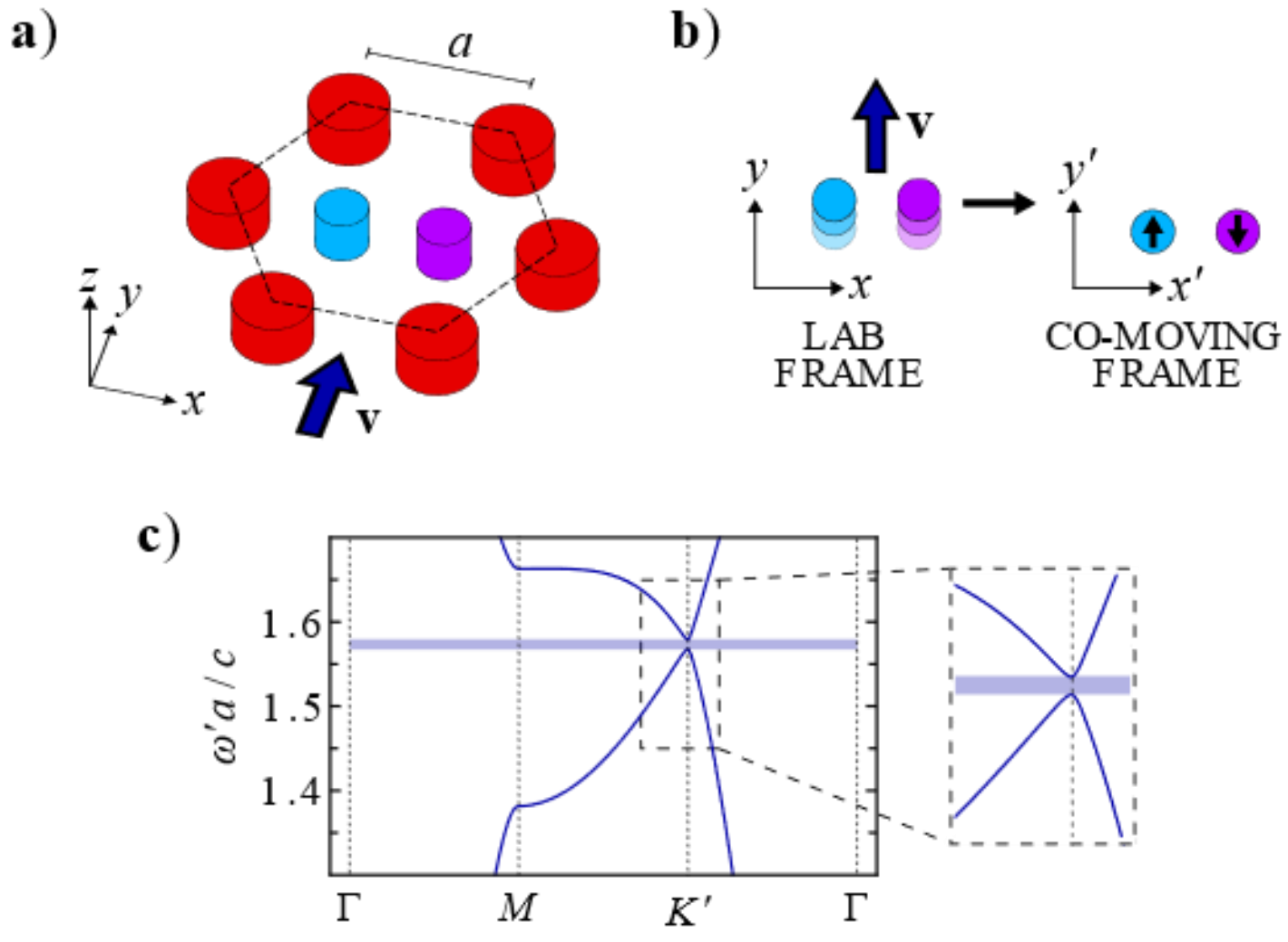

**Fig. S5 a)** Unit cell of a honeycomb array of scatterers (red elements with radius $R_r=0.2a$) with the same refractive index as the vacuum background: $\varepsilon=10\varepsilon_0$, $\mu=0.1\mu_0$. The isorefractive structure is perturbed with two additional elements of radius $R_b=0.25a$ (bluish colors) with parameters $\varepsilon=\varepsilon_0$, $\mu=1.5\mu_0$ (light blue element) and $\varepsilon=\varepsilon_0$, $\mu=0.5\mu_0$ (dark blue element). The modulation velocity is $\mathbf{v}=v\hat{\mathbf{y}}$; in the lab frame, the crystal is Lorentz contracted. **b)** Synthetic magnetic potential $\mathbf{A}$ (black arrows) in the Lorentz co-moving frame implemented by the blue isotropic elements. **c)** Band structure in the Lorentz co-moving frame for a modulation speed $v=0.3c$. The topological charge is $\mathcal{C}_{\text{gap}}=-1$.

Interestingly, the "ring" with two scatterers can be implemented using only isotropic materials. In fact, the vector potential in the isotropic case reduces to $\mathbf{A}\approx-\left(n^2-1\right)\mathbf{v}/c$ and hence it is possible to engineer the *orientation* $\pm\hat{\mathbf{y}}$ simply using materials with a refractive index greater and smaller than the background material (see Fig. S5b). To this end, we use scatterers with the material parameters $\varepsilon = \varepsilon_0$, $\mu = (1 \pm 0.5)\mu_0$ (bluish elements in Fig. S5a). Similar to the main text, as the red elements and the background are isorefractive, only the blue scatterers contribute to the synthetic magnetic potential. Figure S5c depicts the band structure in the Lorentz co-moving frame for $v = 0.3c$. The blue elements are separated by $0.6a$. A detailed analysis shows that the band gap shaded in blue has topological charge $\mathcal{C}_{\text{gap}} = -1$, in agreement with the more complex ring with $N = 3$ elements considered in the main text.

The band gap obtained with the $N = 2$ ring is rather narrow and hence it is more sensitive to perturbations. It is possible to engineer more robust topological phases by considering better approximations of the anisotropic ring. This is illustrated in the final example (Fig. S6) which considers an approximation of the ring using $N = 6$ elements. Now, each element of the ring is an anisotropic material with the permeability main axes controlled by $\varphi_{\mu,n} = \frac{\pi}{2} + \frac{\varphi_n}{2}$ with $\varphi_n = -\frac{\pi}{2} + (n-1)\frac{2\pi}{6}$ and $n = 1,\ldots 6$. Here, we use circular scatterers merely to simplify the numerical modelling.

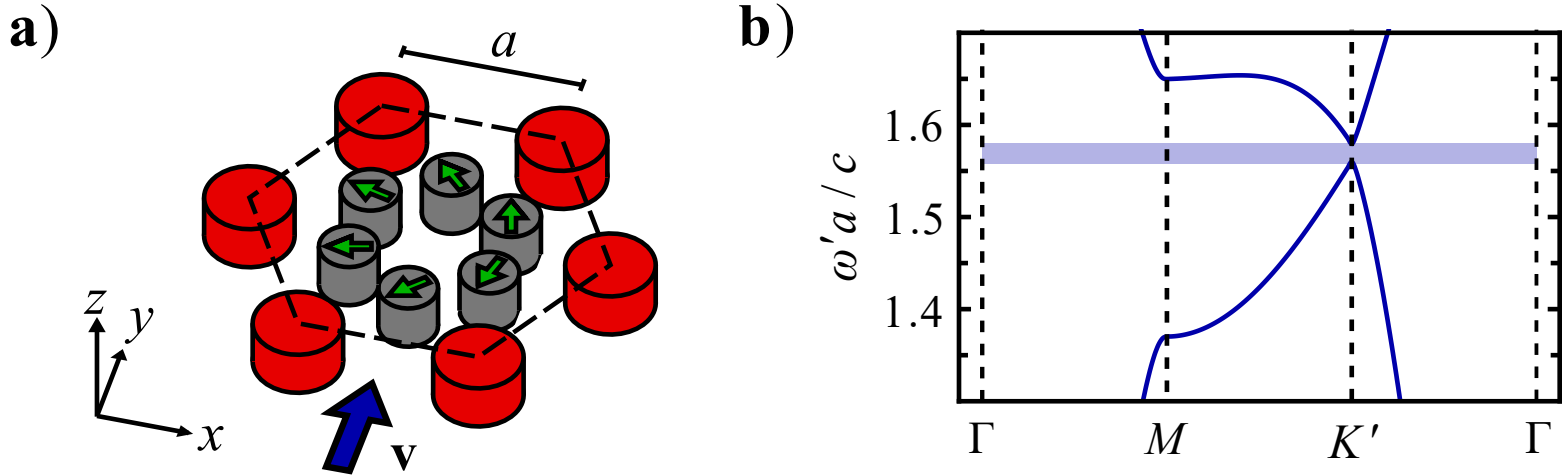

**Fig. S6 a)** Unit cell of a spacetime crystal constructed from an isorefractive system (air background plus the honeycomb array of isotropic red elements with $\varepsilon = 10\varepsilon_0$ and $\mu = 0.1\mu_0$) perturbed with the gray anisotropic elements. The elements in gray provide a discrete realization of the anisotropic ring (Fig. 2bi) with six elements. They have an anisotropic response characterized by the detuning parameter $\delta_{\mu} = 0.5$ and the permeability main axes $\hat{\mathbf{e}}_1$ are represented by the green arrows. The red and gray elements have circular cross-sections with radii $R_r = 0.2a$ and $R_g = 0.17a$, respectively. All the dimensions are specified in the Lorentz co-moving frame. **c)** Band structure in the Lorentz co-moving frame for a modulation speed $v = 0.2c$ along the *y*-direction. The band gap is characterized by $\mathcal{C}_{\text{gap}} = -1$.

Figure S6 confirms that the band gap of the crystal based on the $N = 6$ approximation of the anisotropic ring is approximately $35\%$ wider than the gap obtained with the $N = 3$ approximation. As expected, the topological charge of the band gap remains $\mathcal{C}_{\text{gap}} = -1$, confirming that the topology does not change with $N$. The topological band diagram is qualitatively similar to that in Fig. 3 of the main text (not shown here).